\documentclass{ar-1col-S2O}
\usepackage[numbers]{natbib}
\usepackage{url}
\usepackage{color}
\jname{Ann.\ Rev.\ Nucl.\ Part.\ Sci.}
\jvol{ARNP}
\jyear{2023}
\doi{10.1146/????}

\newcommand{\kmsecMpc}{{km~sec$^{-1}$~Mpc$^{-1}$}}
\newcommand{\gtrsim}{\raisebox{-0.13cm}{~\shortstack{$>$ \\[-0.07cm]
      $\sim$}}~}
\newcommand{\lesssim}{\raisebox{-0.13cm}{~\shortstack{$<$ \\[-0.07cm]
      $\sim$}}~}

\begin{document}

\def\d{{\mathrm{d}}}

\markboth{Kamionkowski and Riess}{$H_0$ and EDE}

\title{The Hubble Tension and Early Dark Energy}

\author{Marc Kamionkowski$^1$ and Adam G.\ Riess$^{1,2}$
\affil{$^1$William H.\ Miller III Department of Physics and
Astronomy, Johns Hopkins University, 3400 N.\ Charles St.,
Baltimore, Maryland, 21218, USA}
\affil{$^2$Space Telescope Science Institute, 3700 San Martin
Dr., Baltimore, MD 21218, USA}}

\begin{abstract}
Over the past decade, the disparity between
the value of the cosmic expansion rate directly determined from
measurements of distance and redshift or instead from the standard $\Lambda$CDM cosmological model calibrated by measurements from the early Universe, has grown to a level of significance requiring a solution.  Proposed systematic errors are not supported by the breadth of available data (and ``unknown errors" untestable by lack of definition). Simple theoretical
explanations for this ``Hubble tension'' that are consistent with the majority of the data have been surprisingly
hard to come by, but in recent years, attention has focused
increasingly on models that alter the early or pre-recombination physics of $\Lambda$CDM as the most feasible.
Here, we describe the nature of this tension,
emphasizing recent developments on the observational side.  We
then explain why early-Universe solutions are currently favored and
the constraints that any such model must satisfy.  We discuss one workable example, early dark energy, and describe how it
can be tested with future measurements.  Given an assortment of
more extended recent reviews on specific aspects of the problem,
the discussion is intended to be fairly general and
understandable to a broad audience.
\end{abstract}

\maketitle

\tableofcontents

\section{INTRODUCTION}

In his ``Chronology of the History of Science and Discovery''
\cite{asimov}, Isaac Asimov identifies Hubble's discovery of the
cosmic expansion as one of the two defining events of 20th-century science (the other being the discovery of the structure of DNA).  Interestingly enough, though, the value Hubble
inferred for the expansion rate (the Hubble constant)---the
ratio of the recessional velocity to distance for the galaxies
he observed---turned out to be too high by an order of magnitude, a rate providing less than 2 billion years for the Universe to
have grown to its present size, far smaller than the age of the Earth!  This first ``Hubble tension'', defined here as any discrepancy between the locally measured and cosmologically inferred expansion rate, was resolved with the
discovery of two generations of stars and a consequence of prior measurements intermingling the two.  

Determination of the Hubble constant has been a central aim of
cosmology ever since, with measurements differing by almost a factor of two as late as the early 1990s
before reaching a celebrated result of $H_0=72 \pm 8$
km~sec$^{-1}$~Mpc$^{-1}$ \cite{Freedman:2001} a 10\% state-of-the-art precision by the new millennium, through use of
the Hubble Space Telescope to resolve Cepheid variables in distant galaxies (later recalibrated to $74 \pm 2$ \citep{Freedman:2012}).  Coupled with the theoretical expectation of $\Omega_m \sim 1$, the low expansion age implied by these measurements, still a few Gyr younger than the oldest stars, set off another ``Hubble tension'' until the discovery of cosmic acceleration amended the composition and recent expansion history and the age of the Universe grew comfortably higher.  

Cosmology has, however, blossomed over the quarter century since then, resembling high-energy physics experiments, 
with huge data sets, sophisticated analyses, careful attention
to systematic errors, and a successful standard cosmological
model ($\Lambda$CDM) with precisely determined parameters. 
Still, two decades
into the new century, we find ourselves yet again with a third Hubble tension, smaller in scale than prior ones but highly significant, a
$\sim8\%$ discrepancy with $\gtrsim 5\sigma$ confidence.  Each past Hubble tension has taught us something more interesting than the value of a parameter and a new tension provides an opportunity for discovery.  Will this one (again) auger new astrophysics or fundamental physics?

The lower value $H_0=67.4 \pm 0.5$
km~sec$^{-1}$~Mpc$^{-1}$ \cite{Planck:2018vyg} of the Hubble constant is
anchored by measurements of angular temperature and
polarization fluctuations in the cosmic microwave background
(CMB) which calibrate free parameters in the $\Lambda$CDM cosmological model. 
Similar values are obtained from spatial fluctuations
in the galaxy distribution \cite{eBOSS:2020yzd,DES:2021wwk} whose physical scale is calibrated by the CMB, an approach called an ``inverse distance ladder.''
These data map the statistical properties of the distribution of
mass to a spectrum of physical scales set by the $\Lambda$CDM cosmological model.  Chief of these scales is the ``sound horizon,'' the distance a primordial fluctuation can travel at the sound speed in an expanding Universe before its size is frozen when the Universe becomes transparent at $z \sim 1000$.   The Hubble constant is one
of six parameters that are optimized to find agreement between this model and the data.
Absent the CMB, comparing primordial deuterium abundance to BBN predicted by the cosmological model in the early Universe  provides similar results, leading to the useful summary that the most precise but indirect measures of the Hubble constant derived from $\Lambda$CDM as calibrated in the pre-recombination or ``early'' Universe give values in the range of 67--68 \kmsecMpc.  

\begin{figure}[htbp]
\includegraphics[width=\textwidth]{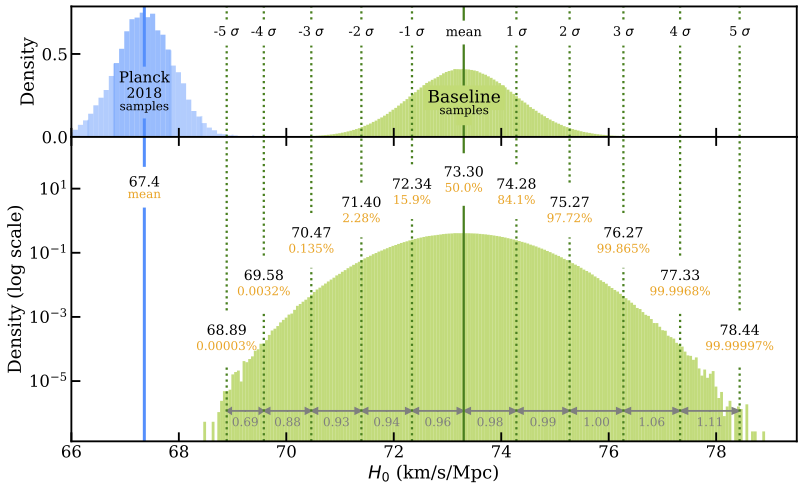}
\caption{Extended MCMC sampling of the posterior for $H_0$ to measure out to the $5 \sigma$  confidence level. The upper panel shows the probability density for the baseline from SH0ES and from the Planck Collaboration et al. (2020) chains. The bottom panel shows the log of the probability density to improve the ability to see the tails. }
\label{fig:local}
\end{figure}

\begin{figure}[htbp]
\includegraphics[width=\textwidth]{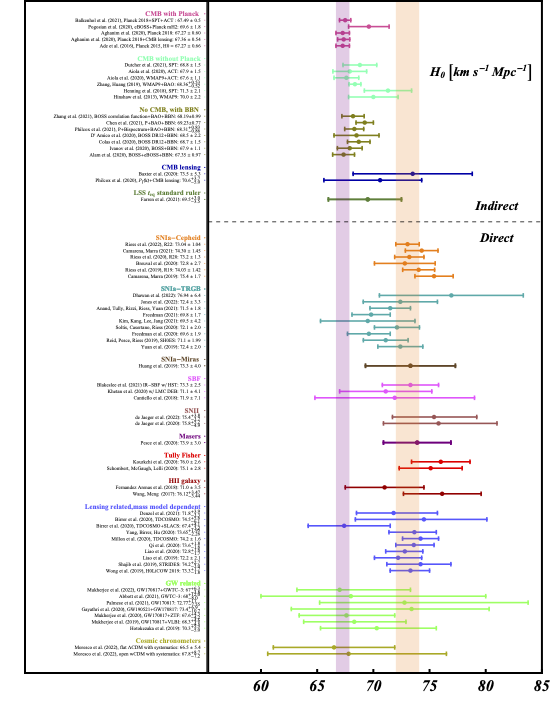}
\caption{68\% confidence-level constraints on H0 from different cosmological probes.  From Ref.~\cite{Abdalla:2022yfr} (based on Refs. [49, 50]).}
\label{fig:whisker}
\end{figure}

A higher range of 70--75 \kmsecMpc\ covers essentially all precise ($\leq$ 5\%), recent ``late'' Universe measurements of the Hubble constant determined ``locally'' or directly---inferred by comparing (as Hubble did) the recessional velocities and distances of galaxies.  
The leading approach in terms of community investment of Hubble Space Telescope time, and most replicated, yields $H_0=73.0\pm1.0$ \kmsecMpc\ 
\cite{Riess:2011yx,Riess:2016jrr,Riess:2021jrx}, near the middle
of the range (see Figures \ref{fig:local} and \ref{fig:whisker}).
Here, the
galaxy distance is inferred from the apparent brightness of a
``standard candle,'' an astronomical source of fixed
luminosity or from the angular size of a ``standard rod'', a source of calculable length. 
While Hubble used Cepheid variables, supergiant stars whose pulsation period correlates with their luminosity, as
standard candles, current local measurements use a variety of
standard candles and rods, but many rely on Type Ia supernovae, a
class of thermonuclear supernovae (stars with degenerate matter which approach the Chandrasekhar mass), to measure deep into the Hubble flow.  We will discuss uncertainties in the measurements in the next Section.

The current Hubble tension has persisted while gaining in significance for nearly a decade, making it hard to ignore.  Well-posed proposals of systematic errors in measurements have been tested and are not supported by the data while non-specific suggestions of measurement ``unknown unknowns'' are unsatisfactory and by definition, untestable.  A viable explanation  
is of interest not just to cosmology, but also to physics.  The standard cosmological
model, assembled in recent decades, is remarkably successful,
but works only with ingredients that involve new physics beyond
general relativity and the $SU(3)\times SU(2) \times U(1)$
Standard Model of elementary-particle
physics.  The Universe is observed on the largest distance
scales to be quite smooth but with small density fluctuations
well described as a realization of a Gaussian random field with
a nearly scale-invariant power spectrum.  The primordial origin
of these perturbations, which have large correlation lengths,
requires new physics (which can be very well described in the
context of inflation) beyond the Standard Model.  The time
evolution of the perturbations in the baryons and in the photons
visible to us as the cosmic microwave background (CMB)---as well as dynamics of galaxies and galaxy clusters in the Universe today---requires some
form of collisionless dark matter which, again, requires new physics.  Moreover, evidence for some
negative-pressure dark energy (e.g., a cosmological constant)
comes from supernova measurements \cite{SupernovaCosmologyProject:1998vns,SupernovaSearchTeam:1998fmf} and the detailed characterization of cosmological perturbations and also necessitates new physics.  

Still, once collisionless dark matter, a cosmological constant, and a nearly scale-invariant spectrum of primordial perturbations are postulated, all of the data we have
on the statistical properties of perturbations in the early and
late Universe can be described by a model that is parameterized
by (i) an overall amplitude for the primordial perturbation
power spectrum; (ii) a power-law index for the power spectrum;
(iii) a baryon density; (iv) a dark-matter density; (v) the
Hubble constant; and (vi) a ``reionization optical depth,''
which quantifies the fraction of CMB photons that are
primordial.

While the Hubble tension looks {\it prima facie} to be a breakdown of $\Lambda$CDM's ability to connect two ends of cosmic time, it does not yield to easily anticipated, new-physics solutions.  In considering the solution space it is important to recognize that $\Lambda$CDM plays two distinct roles in the model-dependent calculation of $H_0$.   First the model is calibrated in its pre-recombination form ($z \gtrsim 1000$) by comparison to primordial measurements which fixes its six free parameters.  Second, $\Lambda$CDM predicts the expansion history between $z \sim 1000$ and $z \sim 0$ which leads to a prediction for $H_0$ with or without additional refinement of the model parameters which may come from comparison to low-redshifts measures of the expansion history.   A late-time solution affecting this extrapolation is attractive but appears less tractable.  If the discrepancy were reversed---with a higher value for $H_0$ coming from the CMB---it could be easily attributed to the nature of dark energy as quintessence-like \cite{Caldwell:2009ix}, where the dark-energy density is slowly decreasing.  However, such an explanation would require, given the larger $H_0$ from local measurements, that dark energy violate the dominant energy condition, something like a relativistic notion of creating energy from the vacuum.  Even if we are willing to allow for such an exotic theoretical
possibility, though, such a model is disfavored by 
galaxy-clustering or high-redshift supernova measurements which prefer a nearly constant $\Lambda$-like dark energy density independent of the primordial measurements.  Neither can the simplest dials on existing dark-matter models be turned to solve the Hubble tension.  Most analyses are done assuming a flat Universe, but allowing for some nonzero curvature actually drives the CMB-inferred $H_0$ even lower.  
Workable solutions, and thus the ones we will focus upon,
can be obtained by modifying $\Lambda$CDM at early times and its early cosmic-expansion history.
One promising way is to postulate some sort of ``early
dark energy'' (EDE) \cite{Karwal:2016vyq,Poulin:2018dzj} that
behaves like a cosmological constant before
matter-radiation equality but then decays away faster than radiation afterwards.  Recent CMB measurements, since EDE was proposed, have improved sensitivity to polarization fluctuations on small angular
scales, and may even favor such models over $\Lambda$CDM.  Still, we must await future CMB
measurements and galaxy surveys to disprove either model.  

A vigorous campaign to develop microphysical EDE models is now under way---at the time of submission of this article, there were on order one new EDE model appearing on arXiv every week.  Most of them repurpose ideas explored earlier in connection with dark energy and/or inflation, but there are some novelties associated with the coincidence between the time that EDE becomes dynamical and the epoch of matter-radiation equality.  The machinery to produce precise model predictions is moreover available, and so are the tools to make detailed comparisons with data sets.  As a result, model building proceeds hand in hand with careful comparisons with ever improving data.  Some of the models suggest entirely new experimental/observational consequences of new EDE physics, which will hopefully bring new avenues to understanding the Hubble tension.  There are also a slew of independent new techniques for local measurements that will further test the results of supernova measurements.

Here we will review the Hubble tension and early
dark energy at a colloquium level.  There have been far too many
developments in measurements and theory in this subject for us to review them in
detail.  We choose, therefore, in our discussion of measurements
to focus primarily on recent developments, current questions,
and future prospects.  The theory discussion will emphasize fundamental issues and model ingredients,
with relevant calculations presented
schematically.  We will then describe the new-physics ingredients for a handful of EDE models.

Fortunately, there are excellent recent and broad reviews
on various aspects of the Hubble tension
\cite{Bernal:2016gxb,Verde:2019ivm,Knox:2019rjx,DiValentino:2021izs,Shah:2021,Efstathiou:2021}
and the various types of models that have been invoked to
explain it \cite{DiValentino:2021izs,Schoneberg:2021qvd}.  Some
recent reviews on (late-time) dark energy
\cite{Frieman:2008sn,Caldwell:2009ix,Weinberg:2013agg} also
cover some overlapping issues in theory and
cosmological-parameter determination.

\section{OBSERVATIONS AND MEASUREMENTS}

\subsection{Defining the Hubble Constant}


The Hubble constant $H_0$ is defined as the constant of proportionality in the relation $cz=H_0 D$ between distance $D$ and redshift $z$  in the limit $z \rightarrow 0$.  For measurements that necessarily involve sources at $z>0$, the linear relation is generalized to
\begin{eqnarray}
D = {cz\over H_0}
\Bigg\{ 1 
- 
\left[1+{q_0\over2}\right] {z} 
+
\left[ 1 + q_0 + {q_0^2\over2} - {j_0\over6}   \right] z^2 
+ O(z^3) \Bigg\},
\nonumber
\label{E:physical}
\end{eqnarray}
which follows from a Taylor expansion of the scale factor
\begin{eqnarray}
a(t)= a_0 \;
\Bigg\{ 1 + H_0 \; (t-t_0) - {1\over2} \; q_0 \; H_0^2 \;(t-t_0)^2 
+{1\over3!}\;  j_0\; H_0^3 \;(t-t_0)^3 
\nonumber
+ O([t-t_0]^4) \Bigg\}.
\end{eqnarray}
with $H(t) = + (da/dt)/a$ the expansion rate at time $t$; $q(t) = - (d^2 a/ dt^2) \left[ H(t) \right]^{-2}/a$ the deceleration parameter; and $j(t) =(d^3 a/ dt^3) \left[ H(t) \right]^{-3}/a$ the jerk parameter, and so on.  For simplicity these relations are defined without curvature but can be generalized with curvature.  The Hubble ``constant'' is then the expansion rate $H_0$=$H(t_0)$ today (time $t_0$). We can determine $H_0$ (and $q_0$, $j_0$, etc.) from measurements of distances and redshifts directly from this definition and independent of the cosmological model.  

Redshifts are easily measured from the change in wavelength of observed atomic transitions (usually emission lines of galaxies) as compared to experimental, laboratory values.
Relative (also called uncalibrated or scale-free) distance measurements at $D>100$ Mpc, where the $\sim 200-300\, {\rm km/sec}$ ``peculiar'' velocities (random velocities relative to the Hubble flow) become negligible, distances are readily obtained from the brightness of Type Ia supernovae which provide for determination of $q_0$ and $j_0$, while $H_0$ drops out of the calculation \cite{Brout:2022}.  The parameters $q_0$ and $j_0$ can also be constrained from other cosmological data.  In either case, the uncertainties in local determinations of $H_0$ from uncertainties in $q_0$ and $j_0$ are irrelevantly small; a change $\Delta q_0$ changes $H_0$ by $O(\Delta q_0$) in \kmsecMpc.  Measurement of $H_0$ then requires {\it absolute} distance measurements; these must ultimately be calibrated by geometry and are harder to come by.  

\subsection{The Local Distance Ladder: Geometry to Cepheids to SN Ia}

 For reasons related to the homogeneity and luminosity of different classes of astronomical objects, the most widely supported route for measuring $H_0$ has been to construct a 3-step distance ladder using geometry to calibrate Cepheid variables followed by SN Ia.  Here we make reference to the specific implementation and most recent iteration of this approach by the SH0ES Team \cite{Riess:2021jrx}.

\begin{marginnote}[]
\entry{Mpc}{Megaparsec, $3\times10^{24}$ cm, roughly, the typical spacing between galaxies}
\entry{SN Ia}{Type IA supernova, a SN from thermonuclear detonation of a white dwarf when it exceeds the Chandrasekhar mass}
\entry{Cepheid}{A Yellow Supergiant star pulsating in the fundamental (radial) mode whose light curve period strongly correlates to its mass and luminosity}
\entry{SH0ES}{Supernovae and H0 for the Equation of State}
\entry{HST}{Hubble Space Telescope}
\entry{WFC3}{Wide Field Camera 3}
\entry{UVIS/IR}{Refers to the two channels on WFC3: UV/visible (200 to 1000 nm) and infrared (800 to 1700 nm)}
\end{marginnote}

Cepheids have been favored as primary distance indicators for more than a century because they are very luminous (100,000 solar luminosities), extremely precise (3\% in distance per source \cite{Riess:2019cxk}), easy to identify due to their periodicity (since Leavitt 1912), and well understood (since Eddington 1917). They are massive, pulsating supergiant stars, overshooting hydrostatic equilibrium due to the $\kappa$ temperature-dependent opacity mechanism \cite{1966ARA&A...4..353C}.  There is a tight coupling between the period of the pulsation (weeks to months), the mass, and the luminosity of these stars, with the latter inferred {\it empirically} from the former from Cepheids at a common distance. They are also the most consistently calibrated standard candle, an important issue for reduction of errors, thanks to use of a single, stable telescope and instrument, HST WFC3 UVIS/IR used for all measurements in local SN Ia hosts and in three geometric calibrators of Cepheid luminosities: the megamaser host NGC4258 \cite{Pesce:2020xfe}, Milky Way parallaxes from the {\it ESA Gaia} mission \cite{Riess:2018byc}, and the LMC \cite{Riess:2019cxk} (via detached eclipsing binaries) with a precision of 1.5\%, 1.0\% and 1.2\%, respectively.   Milky Way parallaxes from the {\it ESA Gaia} mission in particular and through successive iterations, have become the best source of geometric distance measurements, moving the calibration of Cepheids ahead of other stellar distance indicators and even allowing for additional, self-calibration of {\it ESA Gaia} parallax errors \cite{Riess:2020fzl}.    Near-infrared (NIR) observations of Cepheids are used to mitigate the impact of uncertainties related to dust which reduces systematic uncertainties relative to past, optical-only data.  Just recently, a serendipitous early James Webb Space Telescope (JWST) observation of NGC 1365, a nearby galaxy on the Cepheid-supernova calibration path allowed a measurement of the near-infrared Cepheid period-luminosity relation, with JWST's improved resolution \cite{Yuan:2022edy}. The results are consistent with results from HST, but show that future JWST observations will be an important tool in the increasingly precise characterization of Cepheids.

\begin{figure}[htbp]
\includegraphics[width=\textwidth]{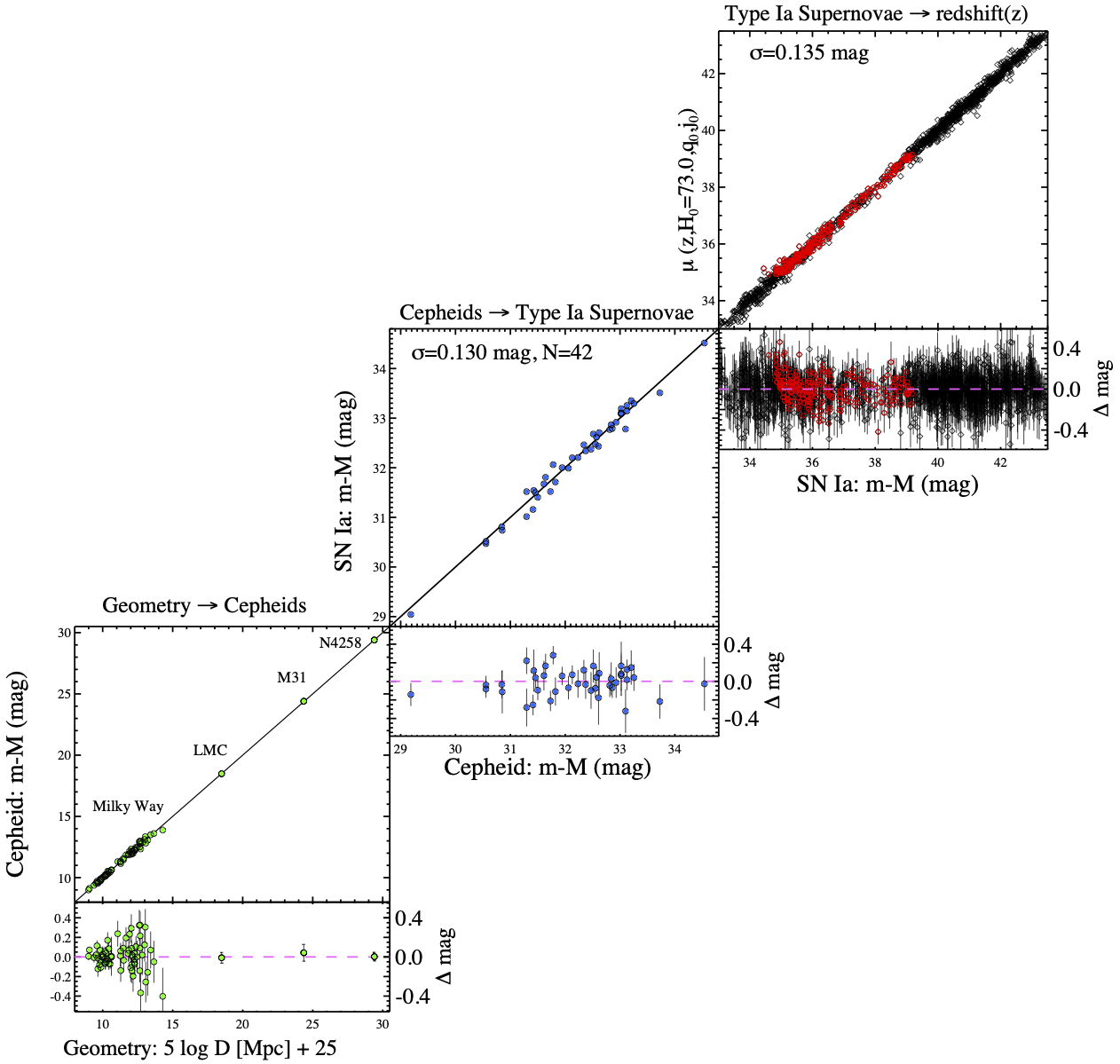}
\caption{The cosmic distance ladder used by SH0ES to infer $H_0$.  The luminosities of nearby Cepheid variables are calibrated to parallaxes.  Supernova luminosities are then calibrated to Cepheid luminosities at larger distances.  The Hubble constant is then inferred from the brightnesses of more distant supernovae.}
\label{fig:3rung}
\end{figure}

SN Ia are rarer than Cepheids (one per galaxy per century vs hundreds per galaxy at any time), hence none have been near enough for a parallax measurement in 400 years, but they are far more luminous (billions of solar luminosities).  Standardizeable to 6\% in distance per source, they have no rival in their ability to witness cosmic expansion.
In the past, the uncertainty in $H_0$ was limited by the rarity of SN Ia whose hosts were in range of Cepheids (at $D \leq 40$ Mpc, about one per year) but better instruments on  HST and persistence has produced a complete sample of forty-two well-observed, prototypical SN Ia from the last four decades (the era of digital photometry).  (The redshifts of these nearby SN Ia hosts do not enter the calculation of $H_0$).  Great efforts have been made to calibrate and standardize these consistently with the thousands of SN Ia in the Hubble flow \cite{Brout:2022,Scolnic:2018} (typically at $0.02 < z < 0.15$), including the modeling of data covariance.  The result is a measurement from the SH0ES and Pantheon+ data of $H_0=73.04 \pm 1.04$ km sec$^{-1}$ Mpc$^{-1}$ including systematic uncertainties; see Figure \ref{fig:3rung}.

\begin{figure}[htbp]
\includegraphics[width=0.9\textwidth]{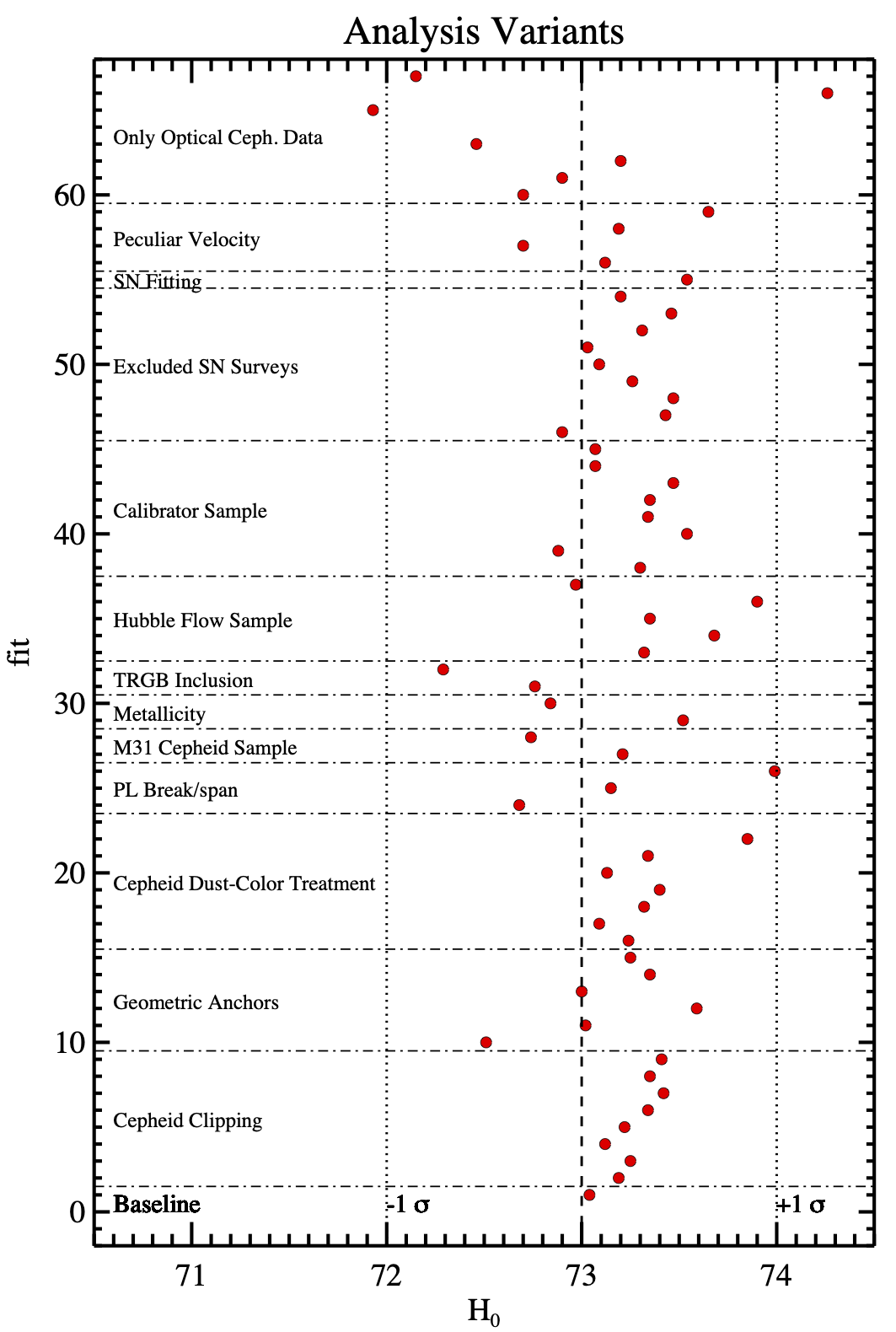}
\caption{Values of $H_0$ inferred by SH0ES under different assumptions and analyses.  The $\pm1\sigma$ vertical dotted lines indicate the statistical error in the baseline result shown at the bottom.  $H_0$ is given in units of \kmsecMpc.}
\label{fig:variants}
\end{figure}

This result has passed a wide range of null tests and has been replicated from the published Cepheid photometry \citep{CSP:2018rag,Feeney:2017sgx} and the Cepheid photometry has been replicated independently \cite{Javanmardi:2021viq}.  Sixty-seven variants of the baseline analysis, see Figure \ref{fig:variants}, demonstrate it is difficult to move the central value below $\sim$ 72.5 or above $\sim$ 73.5 (see Ref.~\cite{Riess:2020fzl} for discussion of uncertainties). The relation between Cepheid metallicity and luminosity, a past source of uncertainty, has been well-calibrated \citep{Breuval:2022}, and due to the breadth of anchors, has little effect on $H_0$ in any case.  Well-posed, experimental challenges to these measurements have been extensively studied and are certain to continue but the present evidence does not support a significant challenge to the conclusions of a highly significant tension.  The comparison to Planck with $\Lambda$CDM yields the strongest evidence for the Hubble tension at 5$\sigma$, or 5.3$\sigma$ including new calibrations from {\it Gaia} clusters \cite{Riess:2022cl}.

\begin{marginnote}[]
\entry{SBFs}{Surface-brightness fluctuations}
\entry{TRGB}{Tip of the red-giant branch}
\entry{AGB}{Asymptotic-giant branch (like red giants, but having burned their helium cores to carbon or heavier elements)}
\entry{SN-II}{Type II supernova, a core-collapse SN distinguished by hydrogen in its spectrum}
\entry{CCHP}{Chicago-Carnegie Hubble Project}
\entry{EDD}{Extragalactic Distance Database}
\end{marginnote}

Nevertheless, it is important to test the individual rungs of this ladder which can be done with other independent distant indicators, e.g., SBFs, TRGB, SN~II, Miras, etc.\ (see figure \ref{fig:whisker}), which presents a composite view of a large sample of local-measurements results).  For reference, to calibrate the nearest SN Ia, a distance indicator needs to reach galaxies at $D>10$ Mpc.  To well-calibrate the Hubble flow, a distance indicator needs to reach $D\sim 100$ Mpc, so that $O(1000\, {\rm km}~{\rm sec}^{-1})$ peculiar velocities are small compared to the Hubble-flow velocity.  To be calibrated by parallax the distance indicator needs to be present in the Milky Way.

\begin{figure}[htbp]
\includegraphics[width=\textwidth]{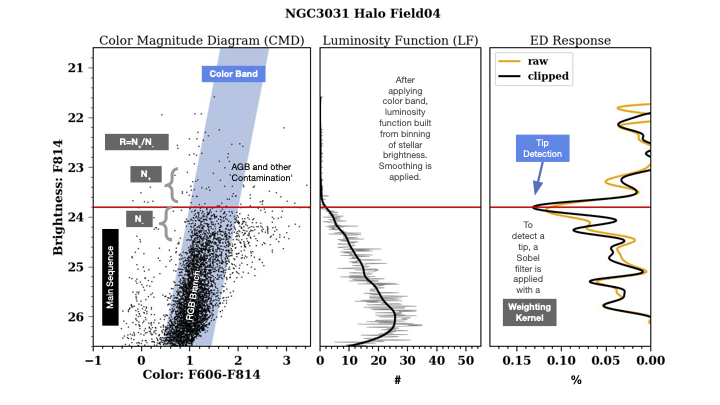}
\caption{This Figure, from Wu et al 2022, illustrates the steps involved in the inference of distances from the tip of the red giant branch and some of the metrics used to judge the quality of the result.  The tip contrast $R$ is the most important of these.}
\label{fig:explainer}
\end{figure}

\subsection{Other Local measurements}
\subsubsection{Tip of the Red Giant Branch (TRGB)}

When a main-sequence star has burned all the hydrogen in its core to helium, the star continues to burn hydrogen in a shell around the core, thus slowly increasing the helium-core mass.  During this time, the star becomes cooler but brighter, a process that continues until the mass, and thus temperature, of the helium core becomes large enough (about 0.5 solar masses) to ignite helium burning to carbon.  The luminosity of the I-band (in the infrared) is seen to be insensitive to the star's metallicity and mass, and so they can be used as a standard candle.
Since the TRGB luminosity is 10 times fainter than long-period Cepheids, it is more limited in distance to $D < 20$ Mpc.  It has, however, been used in lieu of Cepheid variables to calibrate nearby Type Ia supernovae.  

In practice the location of the tip in the color-magnitude diagram of stars often appears ``fuzzy'' due to the presence of AGB stars which have the same color as RGB stars but are both brighter and fainter, reducing the contrast and blurring the location of the tip; see Figure \ref{fig:explainer} for an illustration of the TRGB measurement process.
Varying degrees of contamination of the old, metal-poor halo tip by new star formation or crowding may be partially mitigated by careful selection of regions to analyze \cite{Hoyt:2021} but leave some ambiguity about the location of the tip which is field dependent. Techniques to measure the tip include Sobel edge detection, parameteric luminosity function fitting, and maximum likelihood methods. The TRGB, like Cepheids, are not generally used to measure $H_0$ directly, but contribute to such a measurement by calibrating longer range indicators.

A direct comparison of Cepheids and TRGB and distance measures can be made in seven SN Ia hosts (both deriving calibration from the same geometric distance measure in NGC 4258) and these yield agreement to better than 2\% in the mean \cite{Riess:2020fzl}.  However, literature differences in the determination of $H_0$ involving TRGB are seen to arise from the other two rungs; the (first) calibration rung and a (third) SN Ia rung.  There are several measures of $H_0$ that use TRGB including the Chicago-Carnegie Hubble Project (CCHP) \cite{Freedman:2019jwv} ($70 \pm 2$), the Extragalactic Distance Database EDD \cite{2021AJ....162...80A} ($71.5 \pm 2.0$) which connect to SN Ia, and a measure that calibrates surface brightness fluctuations with TRGB \cite{2021ApJ...911...65B} ($73 \pm 3$).  The main source of differences in the CCHP and EDD TRGB studies come from differences in the determination of the TRGB in NGC 4258 which differ by 1.5 \kmsecMpc\ (in $H_0$).  These groups use different fields in NGC 4258 and more work may elucidate how these fields compare to the fields used to measure TRGB around SN Ia hosts. 
Parallaxes from Gaia of field stars \citep{Li22} or the nearby, massive globular cluster, $\Omega$ Centauri \citep{Soltis21} may also aid TRGB calibration (see Figure 6 in \citep{Li22}  for a summary of recent TRGB calibrations).
The lower value of $H_0$ from the CCHP study also partially sources from differences in its measurement of the Hubble flow from SN Ia; accounting for local peculiar motions \cite{Peterson:2021hel} and cross-calibration of SN samples \cite{Brownsberger:2021uue} would raise that $H_0$ by $\sim$ 1.5 \kmsecMpc\ (or likewise we find removing these steps in the SH0ES Team measurement lowers that measurement to 71.8 \kmsecMpc).  These are small differences, important to understand to reach a 1\% uncertainty.  However, as seen in
\cite{Abdalla:2022yfr} Figure 2, {\it all} local measures at this precision level are above the 67.4 $\pm 0.5$ \kmsecMpc\ expectation (Planck) while none of the local measures are in tension with each other (at $>$ 1.5 $\sigma$).  Even the lowest value of these measures is above the 4$\sigma$ confidence level of Planck.

\subsubsection{Gravitational waves}

The
gravitational-wave (GW) signal from a merging neutron-star binary
can be used to determine the Hubble parameter \cite{Schutz:1986gp}.  The spindown
frequency and GW amplitude can be used to determine the distance
to the source, as long as the inclination of the binary is
known.  The inclination can in principle be constrained with
knowledge of the GW polarization.  Even without polarization,
the distribution of inclination angles can be averaged over in
multiple events.  This measurement requires, though, a redshift
obtained from an electromagnetic counterpart.

This technique was implemented with the detection
\cite{LIGOScientific:2017vwq,LIGOScientific:2017ync,LIGOScientific:2017zic} of a gravitational-wave signal consistent with a neutron-star--neutron-star or neutron-star--black hole merger at a distance $\sim40$~Mpc, as well as detection of an associated soft gamma-ray burst and a slew of subsequent electromagnetic observations.  This event, GW170817, then provided
a measurement $H_0 = 74^{+16}_{-8}$ \kmsecMpc\ ($68\%$ CL) of the Hubble parameter
\cite{LIGOScientific:2017adf}, but the associated
uncertainties---dominated by that for the inclination
angle---were too large to be of value.  Subsequent detection and
modeling of a late-time radio jet from the merger then constrained the inclination angle and thus improved the
measurement to $H_0 = 70.3^{+5.3}_{-5.0}$ \kmsecMpc.  The error reduction is, however, accompanied possibly by radio-jet modeling uncertainties.  In addition, this event at $D \sim 40$ Mpc ($z \sim 0.01$) was too close to provide a clean measure of the Hubble flow with peculiar velocities producing an 8\% uncertainty.
Given the expected increase over the next decade
in detection range, as well as improved characterization from
multiple observatories, a robust
$\lesssim2\%$ GW measurement of the Hubble parameter may be possible in the coming decade.

It may also be possible to constrain $H_0$ from GW events with{\it out} optical counterparts.  The idea of this ``statistical-siren'' or ``dark-siren'' approach is that if the origin of the GW signal can be localized on the sky, then a catalog of candidate host galaxies can be obtained.  Although $H_0$ cannot be determined from this method, the redshift distribution of the candidate hosts translates to a probability distribution for $H_0$ \cite{Chen:2017rfc}.  By combining results from a large number of such events, the possible values of $H_0$ can be narrowed.  The LIGO/Virgo and Dark Energy Survey (DES) Collaborations applied this technique in Ref.~\cite{DES:2019ccw} to GW170814, a particularly bright and reasonably well localized event, although the $H_0$ constraints from this initial foray are not yet significant.

\subsubsection{Mira variables} 

Mira variables are pulsating, low-mass, intermediate age (AGB) stars with great luminosity, comparable to Cepheids in the near-infrared.  The oxygen-rich subtype has a simple, linear period-luminosity relation.  However, these objects are more challenging to find and measure than Cepheids because their periods are far longer, hundreds to a thousand days, and there are contaminating stars in the same period range (Carbon-rich Miras and long-period-variables) which must be distinguished using light-curve amplitudes and colors.  First attempts to use Miras to supplant Cepheids or TRGB are promising \cite{Huang:2019yhh}, but the sample size is small.

\subsubsection{Surface Brightness Fluctuations (SBF)}

This technique determines distances to galaxies by the  surface-brightness fluctuations that arise from the finite number of stars in the galaxy \cite{1988AJ.....96..807T}.  The electromagnetic energy flux $F$ from a given galaxy is related to its distance $D$ by $F\propto D^2$, and the surface brightness is $\propto D^{-2}$.  If the galaxy is resolved, then there will be pixel-to-pixel variations in the surface brightness, induced by the finite number of stars contributing to the flux in any given pixel, with root-variance $\propto D^{-1}$.    This technique yields distances with precision approaching SN Ia but cannot reach the same distances.  Nevertheless, they offer an important alternative to SN Ia and a recent, state-of-the-art study of SBF using {\it HST} in the NIR from \cite{2021ApJ...911...65B,Garnavich22} yields similar results as SN Ia, with $H_0 \sim 73.3 \pm 2.5$ with the same result whether calibrated by Cepheids or TRGB.

\subsubsection{Masers}

Water masers in Keplerian motion around supermassive black holes in the centers of galaxies can be observed in the radio using VLBI.  By tracking proper motions and accelerations, a purely geometric distance can be measured to the maser host.  However, such objects are rare due to the requirement of edge-on alignment of the inner accretion disk with our line of sight coupled with the need for an optimal density profile of the disk.  The Maser Cosmology Project (MCP) has measured 6 such systems in the Hubble flow reported by Ref.~\cite{Pesce:2020xfe} which yields $H_0=74 \pm 3$, independent of any previously reviewed rungs.  While this approach is limited in precision due to the small samples and limited resolution of the galaxy nucleus, future observations of these same maser hosts with the Event Horizon Telescope \cite{EHT} could yield a dramatic improvement and are highly anticipated.

\subsection{Model-dependent local measurements}

We now review several additional techniques to determine the Hubble constant that we refer to as ``model-dependent'' because they require additional modeling:  For gravitational lensing, it is a model for the lens mass; for ages/aging, it is a model for stellar evolution.

\subsubsection{Strong gravitational lensing}

If a given time-varying cosmological source is multiply lensed
by a massive foreground object,  there will be time delays
between the light curves observed in the different images.
These time delays depend on the angular-diameter distances to
the lens and source, thus providing a route to determination of
$H_0$ \cite{Refsdal:1964nw}.  Quasars have proved to be ideal
targets for this measurement as they are bright, time-variable,
and long-lived \cite{1989A&A...215....1V, Keeton:1996kf,
Schechter:1996fa, Koopmans:2003ha}.
In 2019, the H0liCOW Collaboration reported $H_0 =
73.3^{+1.7}_{-1.8}$ \kmsecMpc from an analysis of 6 strongly
lensed quasars with time delays \cite{Wong:2019kwg}.
One issue in the measurement is the mass-sheet degeneracy
\cite{1985ApJ...289L...1F}: a flat mass distribution in the plane along
the line of sight can add to the time delay without affecting
the image locations and brightnesses.  In practice, the mass
distribution of the lens must be constrained with dynamical
constraints to the lens mass, thus introducing new uncertain
astrophysics into the measurement.  As an illustration of the
possible impact, a subsequent analysis of the
H0liCOW quasars (with a seventh added) relaxing the assumption of a conventional galaxy mass profile for the lenses (NFW or power-law) in one of two different ways obtained values consistent with either end of the $H_0$ discrepancy  
and with larger error bars
\cite{Birrer:2020tax}.  So if lenses share the same mass profiles as local, well-studied elliptical galaxies, lensing conforms with the other local values, and if it does not, the way in which it does not becomes the leading source of uncertainty.
The strong-lensing determination of
$H_0$ is now being advanced by the TDCOSMO Collaboration
\cite{Millon:2019slk} with new systems, new analysis pipelines,
and careful attention to identification and mitigation of
systematic effects and astrophysical uncertainties.  

\begin{figure}[htbp]
\includegraphics[width=\textwidth]{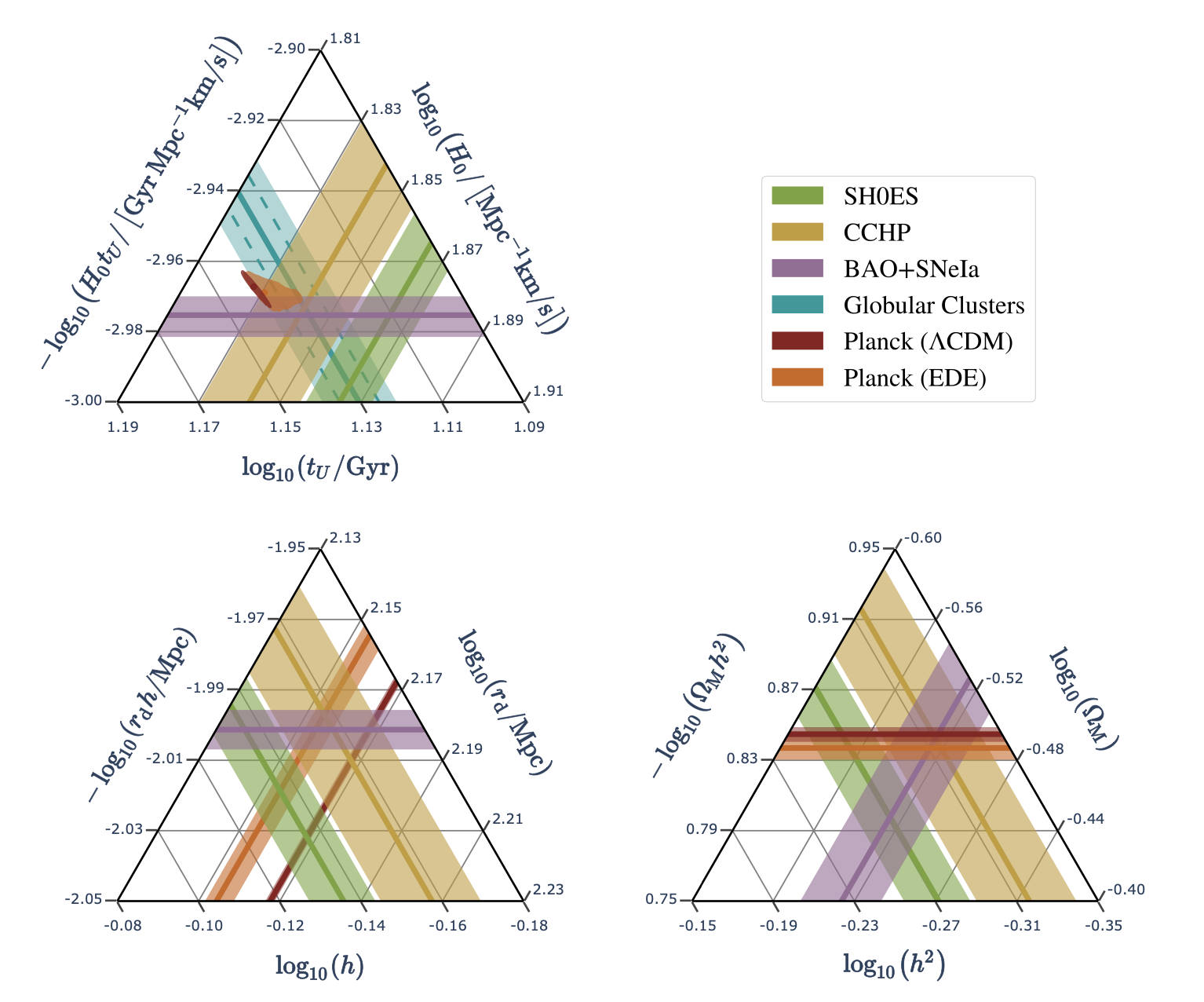}
\caption{The $1\sigma$ allowed parameter regions in triads
     corresponding to the age of the Universe and the Hubble
     constant (upper left); the sound horizon at radiation drag
     and the reduced Hubble constant (bottom left); and the
     total matter density and the square of the reduced Hubble
     constant (bottom right). Note that all points in each
     panel sum to 0, while the ticks in the axes determine
     the direction of equal values for each axis.  From Ref.~\protect\cite{Bernal:2021yli}.}
\label{fig:triangles}
\end{figure}

\begin{marginnote}[]
\entry{$\Lambda$CDM}{The standard cosmological model with dark energy taken to be a cosmological constant}
\entry{wCDM}{The cosmological model with dark energy with an equation-of-state parameter $w$}
\entry{LIGO}{Laser Interferometric Gravitational-Wave Observatory}
\entry{Virgo}{The European gravitational-wave observatory}
\entry{DES}{Dark-Energy Survey}
\end{marginnote}

\subsubsection{Ages and Aging}

The expansion rate $H(z)$ at any given redshift $z$ is
inversely proportional to the ratio of the time interval $\Delta
t$ associated with some given redshift interval $\Delta z$.
Ref.~\cite{Jimenez:2001gg} thus proposed using differential
stellar ages to determine the expansion history, and at low
redshifts, the Hubble parameter $H_0$.  This type of measurement
currently is most precise at a redshift $z\simeq0.45$
\cite{Stern:2009ep,Moresco:2012jh}. This measurement
is subject to stellar-astrophysics uncertainties which have only recently been well-studied by Refs.~\cite{Moresco:2020fbm,Borghi:2021zsr}.
Folding these in constrains $H_0$ to $67.8^{+8.7}_{-7.2}$ and $66.5 \pm 5.4$ \kmsecMpc, respectively, for a generic open wCDM and for a flat $\Lambda$CDM cosmology, results which are not very constraining in the present landscape.

The age of the Universe also
provides a constraint to $H_0$.  In a wCDM cosmology with matter density $\Omega_m$ and dark-energy equation-of-state parameter $w$, it is
\begin{equation}
     t_0 = \frac{1}{H_0} \int_0^\infty \frac{ dz}{(1+z) \sqrt{\Omega_m (1+z)^3 +(1-\Omega_m)^{-3(1+w)}}}.
\end{equation}
The oldest (observable) stars in the Universe are found in low-metallicity (thus formed from nearly primordial gas) globular clusters in the Milky Way halo.  The stars in these globular clusters have nearly uniform metallicities and exhibit a clear main sequence in their color-magnitude diagrams, thus suggesting a uniform burst of star formation.  The age is determined by fitting main-sequence isochrones from stellar-evolution models to the observed color-magnitude diagram, as the main sequence is the best understood phase of stellar evolution.  Recent parallaxes from Gaia have reduced the uncertainties in the stellar luminosity and thus in the inferred ages.  Using ages from 22 Milky Way globular clusters \cite{2017ApJ...838..162O}, Ref.~\cite{Jimenez:2019onw} obtains a
value $H_0= 71.0 \pm 2.7$ \kmsecMpc.  
This provides a
distance-ladder independent measurement and thus complements
other $H_0$ determinations.  The degeneracies with other
cosmological parameters also complement those of CMB and local
measurements, as shown in Figure~\ref{fig:triangles}
\cite{Bernal:2021yli}.  

\section{EARLY-UNIVERSE MEASURES}

\subsection{The sound horizon, the cosmic microwave background, and large-scale structure}
\subsubsection{The early Universe}

The density of the early Universe (``early'' here means within
the first $\sim400,000$ years of the Universe, before the CMB
photons last scattered) was the same to $\lesssim 10^{-5}$
everywhere.  It consisted of photons, baryons ($\sim75\%$
protons by weight, $\sim25\%$ alpha particles, and electrons),
all three neutrino mass eigenstates, and dark matter.  The
cosmological constant (or other form of dark energy) was dynamically insignificant.  Efficient electron-photon
scattering implies that the photons and baryons comprised one
tightly coupled baryon-photon fluid.  The neutrinos were
non-interacting (from a few seconds after the Big Bang) but had
a thermal velocity distribution with a temperature $\sim0.7$
times the photon temperature.  The dark matter is assumed to be
entirely collisionless, an assumption verified by increasingly constraining null searches for dark-matter interactions with baryons, photons, neutrinos, or itself.

The primordial Universe was also populated by adiabatic density
perturbations well described as a realization of a gaussian
random field with power spectrum $P(k) \propto k^{n_s}$ as a
function of wavenumber $k$.  Here, ``adiabatic'' implies that
the fractional density perturbation in each species was
equivalent; i.e., the baryon:photon:DM:neutrino ratio was the
same everywhere.  The scalar spectral index $n_s$ is determined
empirically to be $n_s\simeq0.96$.  The characteristics of this
density field---i.e., $n_s$ is close to, but not precisely equal
to, unity; the adiabaticity; and gaussianity---are all consistent with the simplest single-field slow-roll models of
inflation.  The details of inflation (or even whether it ever occurred) are not relevant to the Hubble tension---we can simply take the flat Universe and nearly scale-invariant spectrum of adiabatic perturbations as empirical facts.

\subsubsection{The sound horizon in the baryon-photon fluid and baryon acoustic oscillations
}
Consider a Dirac-delta-function adiabatic overdensity of matter at some particular point in an otherwise perfectly
homogeneous early Universe.  The pressure in
the baryon-photon fluid associated with this overdensity drives a shock wave that expands at
the sound speed $c_s$ of the baryon-photon fluid (see Figure
10 in Ref.~\cite{Weinberg:2013agg}).  Since the
energy density at these early times is dominated by the photons, this sound speed $c_s$ is just a bit
smaller than $c/\sqrt{3}$, where $c$ is the speed of light.
When photons and baryons decouple, at a time $t_{\rm ls} \simeq
400,000$ years after
the Big Bang (when the plasma temperature falls to $T\lesssim$eV
allowing electrons to combine with nuclei), the shock-induced
overdensity in the baryon-photon fluid has a radius $\sim c_s
t_{\rm ls}$.

The solution to the fluid equations with this
Dirac-delta-function initial condition provides the Green's function for the time evolution of primordial perturbations.  When it is convolved with the primordial mass distribution, a realization of a random field, it provides the two-point correlation function at some later time. Thus, the two-point
correlation function for the baryon density---and thus the
total-matter density, given that baryons constitute
$\sim1/5$ of the total matter density---has a bump at a
comoving distance given by the sound horizon at matter-radiation equality.  
This bump shows up in the galaxy autocorrelation function at a distance $\sim150$ Mpc.  The relatively sharp feature in configuration space then gives rise to oscillatory structures in the Fourier domain.  These are the celebrated baryon acoustic oscillations in the matter power spectrum.

\subsubsection{$H_0$ from acoustic oscillations in the CMB power spectrum}

These oscillations also appear as the acoustic peaks in the angular power spectrum $C_l$ of the cosmic microwave background, since the photon density traces the baryon density at the time, $\sim400,000$ years after the Big Bang, the photons are released.  In 1995, Ref.~\cite{Jungman:1995bz} argued that measurement of these acoustic peaks could be used to  determine the Hubble constant, along with the values of other cosmological parameters, by comparing theoretical calculations of $C_l$ with measurements.  The way that the Hubble constant comes out of this black box can be understood heuristically, however.

The multipole moment $l_s$ of the first acoustic peak determines the angle subtended by the sound horizon at the surface of last scatter, given the correspondence $\ell_s \simeq 2/\theta_s$ between the angular variation $\theta$ of a spherical harmonic of multipole $l$.
The angle subtended by the sound horizon is $\theta_s=
r_s/D_A$, where $D_A$ is the angular-diameter distance to the CMB surface of last scatter, and $r_s \sim c_s t_{\rm dec}$ is the sound horizon.  The parameter $\theta_s=(1.04109\pm 0.00030)\times 10^{-2}$ is the most precisely
determined parameter extracted from CMB measurements, determined to roughly one part in $10^4$.

More precisely, the sound horizon is obtained by integrating the sound speed $c_s(t)$ over time from the Big Bang to recombination.  The comoving sound horizon can be represented by an integral,
\begin{equation}
    r_s = \int_{z_{\rm ls}}^\infty \frac{c_s(z)\, dz}{H(z)} = \frac{c}{\sqrt{3}H_{\rm ls}} \int_{z_{\rm ls}}^\infty \frac{dz} {\left[\rho(z)/\rho(z_{\rm ls}) \right]^{1/2} \left(1 + R \right)^{1/2}},
\end{equation}
over redshift $z$.  Here, $z_{\rm ls}\simeq1080$ is the redshift at which CMB photons last scatter, $c_s(z) = c \left[ 3(1+R) \right]^{-1/2}$ is the sound speed of the photon-baryon fluid, with $R=(3/4)(\omega_b/\omega_\gamma)/(1+z)$, and $\rho(z)$ is the total energy density at redshift $z$.    Here, $\omega_b= \Omega_b h^2$ is the current physical baryon density (today), where $h\equiv H_0/(100\, {\rm km}\, {\rm sec}^{-1}\, {\rm Mpc}^{-1})$ is a dimensionless Hubble constant.  This $\omega_b$ is determined by the higher-peak structure in the CMB power spectrum far more precisely than $\Omega_b$ or $H_0$ separately.  Planck's $\Lambda$CDM value is $\omega_b=0.0224\pm0.0001$.  At last scattering $R\sim 0.5$ and it is smaller at higher redshifts.  And $\omega_\gamma =2.47 \times 10^{-5}$ is the physical photon energy density \cite{Fixsen:1996nj}.  The expansion rate at last scattering is
\begin{equation}
    H_{\rm ls} = 100\,  {\rm km}\, {\rm sec}^{-1}\, {\rm Mpc}^{-1}\,  \omega_r^{1/2} (1+z_{\rm ls})^2 \sqrt{ 1+ \frac{\omega_m}{\omega_r}\frac{1}{1+z_{\rm ls}} },
\label{eqn:Hls}    
\end{equation}
where $\omega_m=\Omega_m h^2$ is the physical nonrelativistic-matter density today---this, again, is fixed fairly precisely by the higher-peak structure in the CMB; Planck's $\Lambda$CDM value is $\omega_m = 0.142\pm 0.001$.  In the standard cosmological model, the early-Universe energy density is $\rho(z) \propto \omega_m (1+z)^3 + \omega_r(1+z)^4$.  The physical radiation density is
\begin{equation}
     \omega_r = \left[ 1 + \frac78 N_{\rm eff} \left( \frac{4}{11} \right)^{4/3} \right]\omega_\gamma,
\end{equation}
where the second term accounts for additional nonrelativistic degrees of freedom. In the standard cosmological model, these include the three neutrino mass eigenstates, and $N_{\rm eff}=3.06$ differs slightly from 3 because of the details of neutrino decoupling \cite{Dodelson:1992km}.

The (comoving) angular-diameter distance to the surface of last scatter is then an integral,
\begin{equation}
    D_A = \frac{c}{H_0} \int_{0}^{z_{\rm ls}}
    \frac{dz}{ \left[\rho(z)/\rho_0 \right]^{1/2}},
\label{eqn:angulardiameterdistance}
\end{equation}
from recombination until the current time $t_0$ when the total energy density is $\rho_0$.   The denominator here is $\rho(z)/\rho_0 = \Omega_m(1+z)^3+(1-\Omega_m)(1+z)^{-3(1+w)}$ in the standard cosmological model, with a dark-energy equation-of-state parameter $w$.  The cosmological constant corresponds to $w=-1$.

From $\theta_s = r_s/D_A$, we infer a Hubble constant,
\begin{equation}
     H_0 = \sqrt{3} H_{\rm ls} \theta_s \frac{  \int_{0}^{z_{\rm ls}}
    dz\, \left[\rho(z)/\rho_0 \right]^{-1/2}   }
     {  \int_{z_{\rm ls}}^\infty dz\, \left[\rho(z)/\rho(z_{\rm ls}) \right]^{-1/2} \left(1 + R \right)^{-1/2}     },
\label{eqn:cmbH0}
\end{equation}
from the CMB.  This is a function of $\omega_b$ through its appearance in $R$.  There is a dependence on $\omega_m$ through its appearance in $\rho(z)/\rho(z_{\rm ls})$ at early times and in $H_{\rm ls}$.  There is also a dependence on $\omega_m$ through the appearance of $\Omega_m=\omega_m/h^2$ in  $\rho(z)$ at late times.  Thus, strictly speaking, equation \ref{eqn:cmbH0} is an implicit equation for $H_0$, given that $h=H_0/(100 {\rm km}~{\rm sec}^{-1}~{\rm Mpc}^{-1})$ appears and in the right-hand side.  There is a dependence on $\omega_\gamma$ in the $R$ and in the expression for $\omega_r$, which also depends on $N_{\rm eff}$.  The dependence on $\omega_r$ comes about in $H_{\rm ls}$ and in $\rho(z)$ at early times. The redshift $z_{\rm ls}\simeq1080$ of the last-scattering surface corresponds to the time when the rate for a photon to Thomson scatter from free electrons---which are becoming scarce as they become bound into hydrogen atoms---becomes smaller than the expansion rate.  There is some dependence of $z_{\rm ls}$ on $\omega_b$ and $\omega_m$ that is taken into account in detailed analyses but is too small to be relevant for the Hubble tension.

In practice, all of the unknown cosmological parameters are determined simultaneously by fitting precise numerical calculations of CMB power spectra to data.  Still, $\omega_m$ and $\omega_b$ are determined primarily by characteristics in the CMB power spectrum such as the Silk damping at higher $l$ and the relative heights of the even- and odd-numbered peaks.  The Hubble constant then follows from equation~\ref{eqn:cmbH0}.  Through numerical differentiation of this expression, it can be found that the Hubble constant varies as $(\Delta H_/H_0) \simeq 0.1\,(\Delta\omega_b/\omega_b)$ for small changes $\Delta\omega_b$ to the baryon density (holding all other parameters fixed) and with $\omega_m$ as  $(\Delta H_/H_0) \simeq -0.77\,(\Delta\omega_m/\omega_m)$ (keeping $\Omega_m = \omega_m/h^2$).  This equation also illustrates how some simple modifications to the standard assumptions might affect the results.  For example, if the number $N_{\rm eff}$ of relativistic degrees of freedom is increased, then the radiation density $\omega_r$ is accordingly increased leading to a higher $H_{\rm ls}$ and thus a higher $H_0$.   Alternatively, if we take $w<-1$, then the integral in the numerator of equation~\ref{eqn:cmbH0} is increased, thus leading to a higher $H_0$.

\subsubsection{CMB results}

The first effort to determine $H_0$ from the CMB was in 2000 \cite{Boomerang:2000jdg}, but the results were not constraining because these initial measurements lacked enough information about the higher peaks in the CMB power spectrum to fix $\omega_m$ and $\omega_b$.  This was accomplished with NASA's Wilkinson Anisotropy Probe (WMAP), which arrived at $H_0=69.3
\pm 0.8$ \kmsecMpc\ \cite{WMAP:2012fli} for their final mission value for the Hubble parameter (improving upon their first-year result, $H_0=73
\pm 5$ \kmsecMpc\ \cite{WMAP:2003elm}).\footnote{Incidentally, WMAP's $\sim1\%$ measurement of $H_0$ improved upon the $\sim10\%$ forecast in Ref.~\cite{Jungman:1995bz} because WMAP's capabilities turned out to be better than anticipated in that work, but also because the acoustic-peak amplitudes turned out to be higher than expected with 1995 best-fit cosmological parameters.}\  Subsequently, the European Space Agency's
Planck satellite \cite{Planck:2018vyg} provided power spectra to multipole
moments $\ell\sim2500$, as opposed to $\ell\sim 800$ from WMAP, finding $H_0=67.4 \pm 0.5$
km~sec$^{-1}$~Mpc$^{-1}$.  These measurements
have then been complemented at even smaller angular scales by the ACT and SPT Collaborations, which arrive at similar values of $H_0$ with $\lesssim$\kmsecMpc\ errors \cite{ACT:2020gnv,SPT-3G:2021eoc}.

\subsubsection{Galaxy surveys and baryon acoustic oscillations}

The sound horizon appears as a bump in the galaxy autocorrelation function at a distance scale $\sim150$ Mpc.  In a galaxy-redshift survey, galaxy locations are parameterized by their position on the sky and by their redshift $z$, a proxy for the line-of-sight distance in the limit that peculiar velocities can be neglected.  A pair of galaxies at similar redshift and some fixed angular separation have a physical separation proportional to the angular-diameter distance $D_A(z)$, which is inversely proportional to $H_0$ and has a dependence on $\Omega_m$; cf., equation \ref{E:physical}. A pair of galaxies  along a given line of sight separated in redshift by $\Delta z$ have a physical separation inversely proportional to the expansion rate $H(z)= H_0 \sqrt{\Omega_m(1+z)^3+(1-\Omega)}$.  To provide some indication of the state of the art, the transverse and radial BAO scales were measured in BOSS to $\sim1.6\%$ and $\sim2.7\%$ in redshift bins of width $\Delta z\sim 0.25$ \cite{BOSS:2016apd}.  The degeneracy between $H_0$ and $\Omega_m$ in $H(z)$ or $D_A(z)$ is different at high and low redshifts, and so can be broken by combining BAO measurements at different redshifts (e.g., Figure\ 5 in Ref.~\cite{eBOSS:2020yzd}), and the BAO measurements now span the range $0.15 \lesssim z \lesssim 3$. Using the sound horizon inferred either from the CMB or from the value obtained by fixing the baryon density from big-bang nucleosynthesis then allows a determination of the Hubble parameter with a similar error.  In practice, galaxy-survey analyses typically add to this ``pure-BAO'' measurement information from the correlation-function shape and its time evolution, and then combine with constraints to cosmological parameters from the CMB, weak gravitational lensing of galaxies or the CMB, or other measurements (see, e.g., Figure\ 20 in Ref.~\cite{DES:2021wwk} for a comparison of the constraints derived under various assumptions).  Currently such BAO+ measurements provide (assuming a sound horizon determined from the CMB) $H_0$ values consistent with the CMB value and with errors $\lesssim$ \kmsecMpc\ \cite{DES:2021wwk,eBOSS:2020yzd}.

\subsubsection{Distance scale of matter-radiation equality}

The Hubble parameter can also constrained by the wavenumber $k_{\rm eq}$
of matter-radiation equality obtained from galaxy surveys.  The
primordial linear-theory matter power spectrum $P(k)$ transitions
from its large-wavelength ($k\to 0$) behavior $P(k) \propto
k^{n_s}$ (with $n_s\simeq 0.96$ the scalar spectral index) to
$P(k) \propto k^{n_s-3}$ at $k\to\infty$ at a wavenumber $k_{\rm
eq}$ corresponding to the mode that enters the horizon at
matter-radiation equality.  Given the rough coincidence between
this distance scale and those corresponding to the sound
horizon, the BAO wiggles in the power spectrum must be modeled
out.  Once they have been subtracted, though, the technique
provides a sound-horizon--independent measurement of the Hubble
parameter. This is the idea behind the ShapeFit algorithm \cite{Brieden:2021cfg} which in a preliminary application to BOSS data finds a low $H_0$. It is also the approach in Ref.~\cite{Philcox:2022sgj} which obtains a
value $H_0=64.8^{+2.2}_{-2.5}$ \kmsecMpc.  However, this
measurement assumes the standard $\Lambda$CDM power spectrum.
If, however, the early expansion history is changed in a manner
suggested by early dark energy, then, as discussed below, the Hubble parameter inferred from this
measurement is raised to a value consistent with the local SH0ES measurement
\cite{Smith:2022iax}.

\section{THEORY AND MODELS}

\subsection{Early- versus late-time solutions}

Barring a combination of systematics that address multiple types of observations, the Hubble tension implies
some new physics
beyond the ingredients (collisionless dark matter, a
cosmological constant, and Standard Model interactions for
baryons/photons) found in the standard cosmological model.
Given that local measurements of the Hubble constant are fairly
straightforward, the aim of most solutions to the Hubble tension
is to introduce new physics that increases the value of $H_0$
inferred from the CMB.

These solutions are typically categorized as ``late time'' or
``early time,'' a classification that can be understood from
equation~\ref{eqn:angulardiameterdistance}.  Late-time solutions
postulate that the energy density in the post-recombination
Universe is smaller than in the standard model, holding the
current density fixed: i.e., $\rho(z)/\rho_0(z) \leq \left[
\rho(z)/\rho_0(z) \right]_{\rm standard}$; cf. equation~\ref{eqn:cmbH0}.  This then increases
the comoving distance to the surface of last scatter and thus
leads to a larger $H_0$.  Early-time solutions postulate that
the energy density is somehow increased before recombination so
that the sound horizon at recombination is decreased. We will also discuss models that decrease the sound horizon by changing the
physics of the baryon-photon fluid.

\subsubsection{Late-time solutions}

Given the plethora of models and the continued inventiveness of
theorists, care should be taken in making blanket statements.
Still, there are theoretical reasons that make late-time
solutions unpalatable and empirical constraints that make them
elusive.  A late-time solution requires that the energy density
at times between decoupling and now is smaller than that in the
standard model, but {\it keeping the energy density today
fixed} \cite{Poulin:2018zxs}.  Given that the scaling of the
radiation and matter
densities with redshift is known, this requires some exotic
matter whose energy density {\it increases} with time.  This is
most easily accomplished by postulating that the cosmological
constant is a phantom field \cite{Caldwell:1999ew}, a fluid with
an equation-of-state parameter $w=p/\rho < -1$, where $p$ and
$\rho$ are here the dark-energy pressure and energy density.
This, however, implies a fluid that violates the strong energy condition; i.e., it effectively creates energy out of nowhere.  This seems strange, but is this what the Hubble tension is telling us? Even if we are willing to accept a violation
of the strong energy condition, though, such models are difficult to
reconcile with the sound horizon seen in the galaxy correlation function  \cite{Efstathiou:2021ocp,Keeley:2022ojz}.  They are also difficult to reconcile with constraints to the equation-of-state parameter $w$ inferred recently from SNe Ia at high redshifts \citep{Brout:2022}.

\subsection{Early dark energy}

\begin{figure}[htbp]
\includegraphics[width=0.9\textwidth]{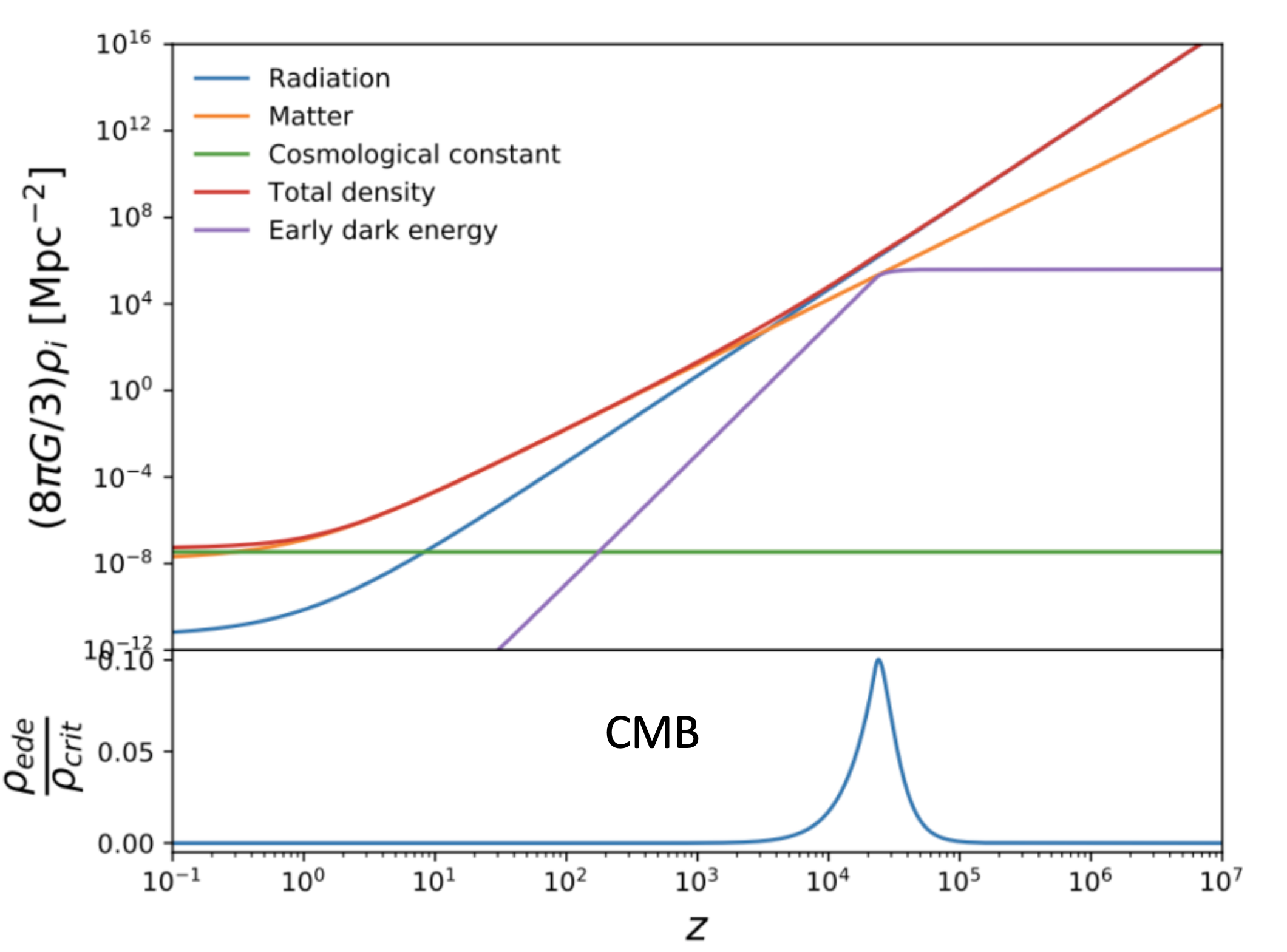}
\caption{The evolution of the energy densities of radiation, nonrelativistic matter (baryons and cold dark matter, and the cosmological constant as a function of redshift (so time increases to the left, with the Big Bang far off to the right and today off to the left.)  Also shown is the energy density postulated for early dark energy (EDE).  The bottom panel shows the fractional contribution of EDE to the total energy density.  The EDE curves are schematic---the key point is that it contributes $\sim10\%$ a bit before recombination but is otherwise dynamically unimportant.  Figure courtesy T.\ Karwal.}
\label{fig:EDEdensities}
\end{figure}

The basic idea behind early dark energy (EDE) is to postulate
some exotic fluid that contributes $\sim10\%$ of the total
energy density of the Universe briefly before recombination and
then has an energy density that decays faster than radiation at
late time, so that it leaves the late evolution of the Universe
unchanged.  This increases $\rho(z)/\rho(z_{\rm ls})$ in the denominator of equation~\ref{eqn:cmbH0}, thus leading to a higher $H_0$.  Although the basic idea is simple, specific models
are highly constrained by the very well measured structure of
the high-$\ell$ peaks in the CMB power spectra.   Fourier modes of the
density field that correspond to the highest multipole moments
($\ell\sim3000$) probed by current measurements entered the
cosmological horizon and became dynamical at a redshift
$z\sim 10^6$, when the Universe was only $\sim$yr old.  The
measured CMB power spectrum thus constrains the expansion
history to far earlier times that the time of last scattering.  Moreover, a fluid with a density
that evolves with time implies, for a
relativistically-invariant theory, the possibility of spatial
fluctuations in the EDE energy density.  This, along with the
already complicated interplay between baryon-photon acoustic
waves, dark matter, neutrinos, and the gravitational field,
implies that any physical model for EDE will be highly
constrained.

\begin{figure}[htbp]
\includegraphics[width=\textwidth]{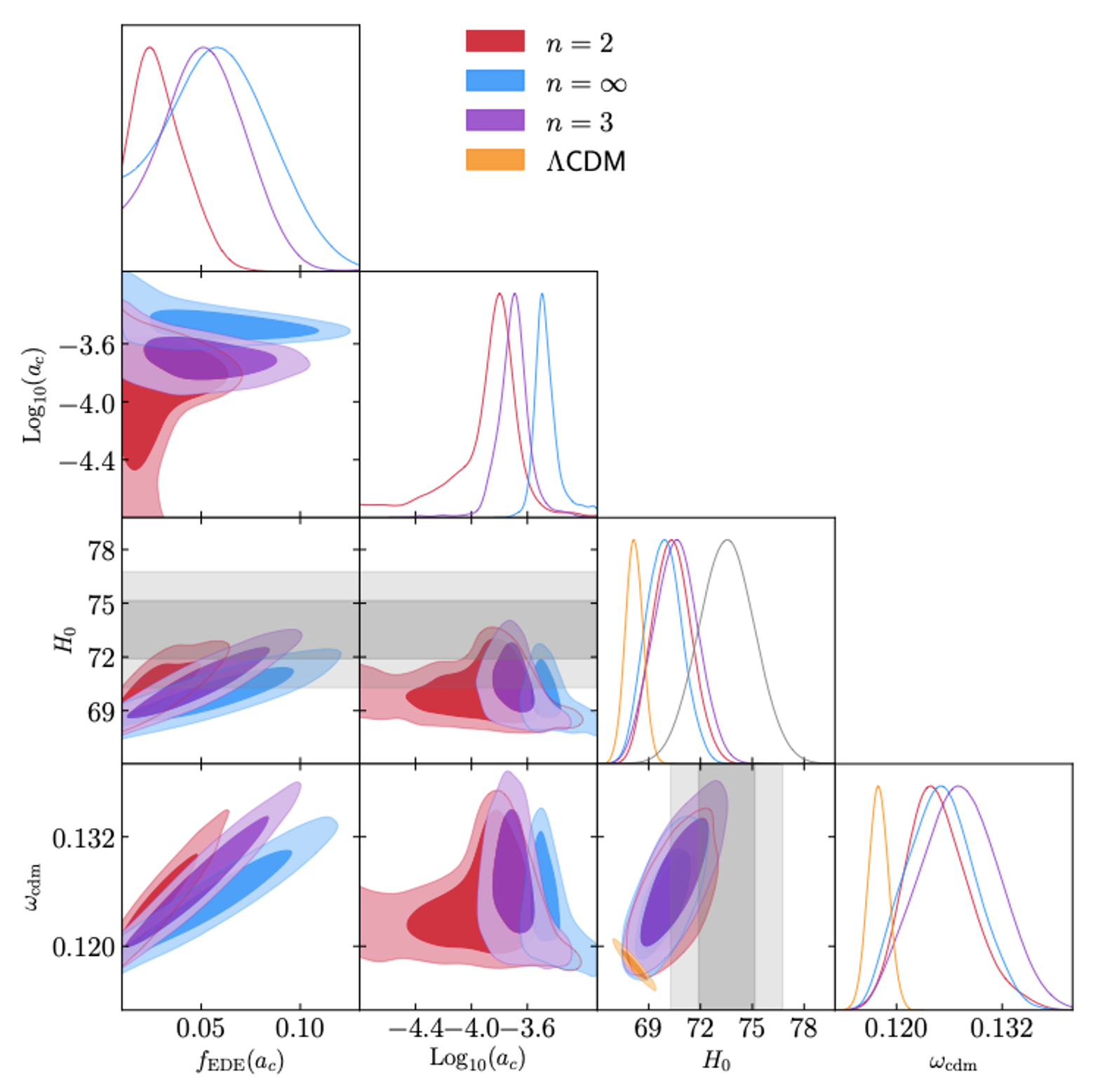}
\caption{Likelihood contours from Ref.~\cite{Poulin:2018cxd} for $\Lambda$CDM and EDE in a subset of the 9-dimensional parameter space.  These are likelihoods for the model given data from CMB and large-scale structure measurements as of late 2018.  The 9 parameters include the 6 $\Lambda$CDM parameters $(\Omega_m,h,A_s,\Omega_b, n_s, \tau)$ along with three EDE parameters: the fractional contribution $f_{\rm ede}(a_c)$ to the total cosmic density at the scale factor $a_c$ (normalized to unity today) at which the this fractional density peaks, and a third parameter which quantifies the sharpness of the transition of EDE from its early-time constant-density behavior to its later decay.  Each panel shows the likelihood contours in a given two-dimensional slice of the full parameter space, after marginalizing over the other 7 parameters.  At the top of each column is the likelihood for the given parameter after marginalizing over all other parameters.  The gray bands and curves indicate the likelihoods for $H_0$ from local measurements   The orange contours show the standard $\Lambda$CDM results and indicate the preference of CMB/LSS data for a low Hubble parameter.  The other three sets of contours describe EDE models in which the EDE decays with scale factor $a$ at late times as $a^{-6n/(n+1)}$.  It is seen that EDE expands the allowed parameter space to regions that overlap the local values of $H_0$.  Although it cannot be inferred from the Figure, it can be checked that the regions of the EDE parameter space that accommodate a larger $H_0$ do so with a reduced-$\chi^2$ similar to that for the best-fit $\Lambda$CDM model.}
\label{fig:poulinplot}
\end{figure}

Even so, it was found Ref.~\cite{Poulin:2018dzj} that physical models of EDE could resolve the Hubble tension.  Figure \ref{fig:poulinplot} shows the results reported there for EDE models in which the EDE density decays with scale factor $a$ at late times as $a^{-6n/(n+1)}$.  As can be seen from the $H_0$ likelihood distribution, the addition of EDE allowed (as of late 2018) for larger values of $H_0$, that overlap the range of values allowed by local measurements, to be inferred from CMB/LSS data.  Moreover, the best-fit EDE model had a similar reduced-$\chi^2$ as the best-fit $\Lambda$CDM model.  The results shown here are for oscillating-field and slow-roll models for EDE, both of which can be described within the required precisions with the generalized-dark-matter formalism. 

\subsubsection{Oscillating-field models}

Oscillating scalar fields underly many of the EDE models that
have been explored.  Here, a scalar field $\phi$ is postulated
with a confining potential \cite{Kamionkowski:2014zda},
\begin{equation}
     V(\phi) = \Lambda_{\rm ede}^4 \left[1-\cos(\phi/f_{\rm ede})
     \right]^n,
\label{eqn:EDEfield}     
\end{equation}
with an energy scale $\Lambda_{\rm ede}\sim$eV so that the
energy density is in the ballpark of that of the Universe before
recombination.  Although this potential was postulated
\cite{Kamionkowski:2014zda} in an {\it ad hoc} fashion,
Ref.~\cite{McDonough:2022pku} argued that it may arise in string
theory.  If the initial field value is $\phi/f \sim \pi$,
then the field is initially frozen and behaves gravitationally
like a cosmological constant.  Later, though, the field begins
to oscillate about its minimum with an equation-of-state
parameter $w_{\rm ede} = (n-1)/(n+1)$ and an energy density that
decays with scale factor $a$ as $\rho_{\rm ede} \propto
a^{-6n/(n+1)}$ \cite{Turner:1983he,Johnson:2008se}.  Thus, for
$n>1$, the EDE energy density decays
faster than that of radiation, as needed.  Another similar
possibility is to simply take the potential to be $V(\phi)
\propto \phi^{2n}$ \cite{Agrawal:2019lmo}.  Oscillating-field
models are parameterized (for fixed
$n$) by the energy density $\Lambda_{\rm ede}$, the decay
constant $f_{\rm ede}$, and the initial field value $\phi_i$.
The power-law index $n$ can be considered as a fourth
parameter.

\subsubsection{Slow-roll models}

Another possibility \cite{Poulin:2018dzj,Lin:2019qug} is a
scalar field with a smooth
``slow-roll'' potential like those considered for inflation or
quintessence.  For example, a potential that asymptotes to
a power law $\phi^n$ for large positive $\phi$ but then asymptotes to zero for
large negative $\phi$ will have an energy density that is
constant at early times and then scales as $\rho \propto a^{-6}$
at late times.   Ref.~\cite{Lin:2019qug} discusses characteristics of EDE potentials and also generalize to models with nontrivial kinetic terms.

\subsubsection{Generalized EDE}

EDE models can also be approached in a slightly more
model-independent way using the generalized dark-matter (GDM) approach
\cite{Hu:1998kj}.  Here, it is noted that communication between
the dark sector (dark matter, dark energy, EDE) and the
gravitational fields occurs in the Einstein equations only
through the stress tensor which can be parameterized in terms of
a fluid density, pressure, sound speed, and viscosity
parameter.  With this approach, the EDE can be described as a
fluid with these parameters.  The approach can be beneficial as
codes that modify the standard cosmological Boltzmann codes
\cite{Bertschinger:1995er,Seljak:1996is,Lewis:1999bs,Lesgourgues:2011re}
to include rapidly oscillating fields are numerically
challenging, given the wide separation between the oscillation
and expansion time scales \cite{Hlozek:2017zzf}.
Ref.~\cite{Poulin:2018dzj} showed how the parameters of the
oscillating-field model can be mapped to GDM parameters, and
Ref.~\cite{Sabla:2022xzj} then mapped EDE-like physics in terms
of a GDM-like parameterization.

\subsubsection{Specific implementations}

One of the curious aspects of EDE models is the coincidence
between the time at which EDE becomes dynamical and the epoch of
matter-radiation equality.  It may be possible to connect these in chameleon models \cite{Karwal:2021vpk} and possibly in some of the alternative-gravity models reviewed below.  This epoch also
coincides with the time at which the heavier neutrino mass
eigenstate(s) become(s) non-relativistic, a realization
capitalized upon in Ref.~\cite{Sakstein:2019fmf}.

Refs.~\cite{Berghaus:2019cls,Berghaus:2022cwf} show that the
friction required for slow-roll models may be induced by
coupling an axion field to a thermal bath in non-abelian
theories.  The idea of interacting thermal baths also plays a
role in the scenario or Ref.~\cite{Aloni:2021eaq}.  Here,
interactions of strongly-coupled radiation are mediated by a
force carrier that becomes nonrelativistic at a temperature
$T\sim$eV.  The mediator then deposits its entropy into
the lighter species thus providing a step in the effective number $N_{\rm eff}$ of relativistic degrees of freedom.  This provides a higher energy density just before recombination while avoiding the problems at high $l$ in the CMB power spectrum associated with an increased $N_{\rm eff}$.

\subsection{Other EDE observables}

If the Hubble tension is solved by EDE, it is natural to inquire
whether the new physics associated with EDE has any other
observable consequences beyond the impact on the expansion
history.  Given the disparity in models, there is no general
consequence, but several possibilities have been discussed.  

\subsubsection{Cosmic birefringence}

If a scalar field $\phi(x)$ is coupled to the electromagnetic field-strength tensor $F_{\mu\nu}(x)$ through a term $[\phi(x)/m_{\rm pl}] \epsilon^{\mu\nu\rho\sigma} F_{\mu\nu} F_{\rho\sigma}$ in the Lagrangian, then time evolution of the scalar field yields a difference in the propagation speeds for right- and left-circularly polarized electromagnetic waves, and thus a frequency-independent rotation of the linear polarization of an electromagnetic wave \cite{Harari:1992ea,Carroll:1989vb,Carroll:1998zi}.  One possibility for such a time-evolving scalar field is quintessence, a candidate for dark energy \cite{Carroll:1998zi,Caldwell:2009ix}.  Another is a slowly rolling scalar field associated with EDE \cite{Capparelli:2019rtn,Murai:2022zur}.  In either case, the rotation angle is $\Delta\phi/m_{\rm {Pl}}$, where $\Delta\phi$ is the change in $\phi$ between the emission and observation of the EM wave (although, strictly speaking, this result is altered if some of the rotation occurs before recombination \cite{Murai:2022zur}).  This ``cosmic birefringence'' (CB) leads to parity-breaking EB and TB (where T, E, and B refer to the temperature and parity-even and parity-odd polarization modes \cite{Kamionkowski:1996ks,Zaldarriaga:1996xe}) correlations in the CMB power spectra \cite{Lue:1998mq,Lepora:1998ix}.
The association of CB with EDE can be
distinguished from late-Universe CB-inducing physics by the
absence of CB observed in late-Universe probes like those
that seek CB in the CMB reionization bump \cite{Liu:2006uh} or
in kinetic-Sunyaev Zeldovich tomography \cite{Hotinli:2022wbk,
Lee:2022udm}. 

\subsubsection{Nonlinear evolution of oscillating EDE field}

If EDE is due to an oscillating field, nonlinear
evolution of the scalar-field perturbations may lead to strong instabilities that then generate nonlinear spatially-inhomogeneous dynamics or soliton-like
structures that then evolve as a subdominant dark-matter component \cite{Smith:2019ihp,Agrawal:2019lmo}.  There is also a possibility that fluctuations in the initial field value may give rise to isocurvature perturbations \cite{Smith:2019ihp}.

\subsubsection{The light horizon}

Changes to the early expansion history will affect the light horizon at decoupling probed by the ``acoustic'' peaks in the
CMB B-mode power spectrum \cite{Jeong:2019zaz}.  These B modes are induced by inflationary gravitational waves, but their amplitude depends on the energy scale of inflation, which is still undetermined.  Observation of these peaks is conceivable with a ground-based Stage-IV CMB experiment if the B-mode amplitude is near the current upper limit.  A space-based experiment, like PICO, may be required, though, if it smaller by an order of magnitude, and the measurement cannot be done if the gravitational-wave amplitude is any smaller.

\subsubsection{Recurrent cosmological-constant-like behavior?}

If the Hubble tension is due to EDE, it also suggests---in
combination with evidence for accelerated expansion today
\cite{SupernovaSearchTeam:1998fmf,SupernovaCosmologyProject:1998vns}
and for inflation in the early Universe---the possibility of
recurrent periods of cosmological-constant--like behavior
throughout the history of the Universe.  This possibility was
anticipated in work on tracking-field models
\cite{Griest:2002cu,Dodelson:2000jtt}, quintessence models in which the  potential is such that the energy density is always a (roughly) fixed fraction of the energy density of the dominant matter component (radiation or matter).  If, however, there are small wiggles added to this potential, then the energy density in the quintessence field can occasionally jump and behave briefly like a cosmological constant before decaying away.  The possibility was also anticipated in a string-axiverse
scenario \cite{Kamionkowski:2014zda}.  Here, in each logarithmic time interval in the Universe's history, there is an axion-like field, with potential as in equation \ref{eqn:EDEfield}, that becomes dynamical.  The initial field value is chosen from random, and if it is sufficiently displaced from the minimum, it can briefly behave dynamically like a cosmological constant.  This scenario also resembles assisted quintessence, explored in connection with EDE in Ref.~\cite{Sabla:2021nfy}.  The idea of recurring cosmological constants motivates the search for other times in cosmic history where the expansion rate can be probed.  Big-bang nucleosynthesis constrains the expansion rate a few minutes after the big bang.  Ref.~\cite{Hill:2018lfx} discussed the possibility to probe the expansion history at redshifts $z\sim17$ with the global 21-cm intensity as measured, for example, by EDGES \cite{Bowman:2018yin}.  Velocity acoustic oscillations---oscillations in the 21-cm angular power spectrum induced by spatial modulation of star formation induced by baryon--dark-matter relative velocities---may also probe the expansion history at similar redshifts \cite{Munoz:2019fkt,Munoz:2019rhi,Sarkar:2022mdz}.  However, there is probably more that can be done along these lines.

\subsection{Other early-Universe solutions}

\subsubsection{Modified gravity}

Modifications to gravity are notoriously difficult:  Since modifications to the Einstein-Hilbert action generically make the scalar degree of freedom (at least, and sometimes also the vector degrees of freedom) in the metric dynamical, there is no obvious way to perturb away from general relativity.  Nevertheless, the accelerated cosmic expansion, the persistent mystery of dark matter, and now gravitational-wave measurements that probe previously inaccessible regions of strong-field gravity have yielded an active marketplace of alternative-gravity theories that can be explored in connection with the Hubble tension \cite{Lin:2018nxe, Braglia:2020auw,
Braglia:2020iik, Ballesteros:2020sik, Ballardini:2020iws,
Zumalacarregui:2020cjh,Abadi:2020hbr}. 

For example, Ref.~\cite{Lin:2018nxe} took a phenomenological approach to modified gravity in which the evolution of cosmological perturbations took on a parameterized departure from those in general relativity.  In relativistic cosmological perturbation theory, there are two scalar potentials $\Phi$ and $\Psi$ (in conformal Newtonian gauge) that generalize the gravitational potential in Newtonian gravity.  The equations that relate these to the energy-density perturbation, pressure, and anisotropic stress are modified in two ways.  First, Newton's constant $G$ in the Fourier-space equations is multiplied by a function $\mu(a,k)$ of the scale factor $a$ and wavenumber $k$.  And second, the the cosmic slip $\Phi-\Psi$ (sourced by the anisotropic stress) is replaced by a $\Phi-\gamma(a,k) \Psi$.  GR is recovered in the limits $\gamma\to1$ and $\mu\to 1$.  A departure of $\mu$ from unity at early times affects the evolution of the gravitational fields so that the phase of the acoustic oscillations is shifted in a way that mimics a shift in the sound horizon.

In conformally coupled gravity, the Ricci scalar $R$ in the Einstein-Hilbert action is multiplied by a function
$(1+\xi \phi^2/m_{\rm Pl}^2)$ of a scalar field $\phi$  with $\xi=-1/6$, so that the scalar is conformally coupled.  The equation of motion for the scalar field is $\ddot \phi +3 H \dot \phi -\xi R \phi=0$.  At early times, during radiation domination, $H^2 \gg R$, and the scalar field is frozen at its initial value. Near the transition to matter domination, when $R$ approaches $H^2$, the field then starts to roll to its minimum (leaving gravity as it is today) and then has an energy density  which decays away as $\rho_\phi \propto a^{-9/2}$ in the matter-dominated era \cite{Abadi:2020hbr}.

\subsubsection{Changing $N_{\rm eff}$}

The idea to reduce the sound horizon by increasing the
early-Universe expansion rate preceded EDE models; it was noted
that an increase in the number $N_{\rm eff}$ of relativistic
degrees of freedom would allow for a larger Hubble constant
\cite{Riess:2016jrr}, but this solution did not provide a good
fit to CMB data, given that an increased $N_{\rm eff}$ affects the higher-$l$ modes in the CMB that probe redshifts up to $z\sim 10^6$.  A fractional increase in $N_{\rm eff}$
in the range of 0.2--0.4 is still allowed and could alleviate the Hubble tension.  Detection of an additional, sterile neutrino would reopen this solution space.

\subsubsection{Changing physics of the baryon-photon fluid}

Models with interacting neutrinos have also been
explored \cite{Kreisch:2019yzn}.  However, they modify the sound
horizon through dynamics of the perturbations rather than
increasing the expansion rate and thus should probably not be
classified as EDE models.  They also run up against laboratory
constraints to neutrino properties \cite{Blinov:2019gcj}.
Primordial magnetic fields have also been suggested as a
solution to the Hubble tension \cite{Jedamzik:2020krr}, but the viability of the idea awaits a more detailed calculation of the evolution of perturbations.

\subsection{Recent results from CMB/LSS data}

There is now a large literature devoted to testing various EDE models with the ever-increasing products of ongoing galaxy surveys and CMB experiments.  The results of Ref.~\cite{Poulin:2018cxd} shown in Figure \ref{fig:poulinplot} have been reproduced, updated, and expanded upon with different data sets. We do not review this work in detail here, as the literature is large and the situation rapidly evolving with new data sets.  

Since the level at which EDE models differ from $\Lambda$CDM are at the $\sim3\sigma$-ish level, small changes in analyses or model assumptions that one might guess were ``below the radar'' can actually affect the conclusions.  For example, the galaxy-clustering constraint to $H_0$ changes by about 0.7 \kmsecMpc\ if the dark energy is assumed to be a cosmological constant or described by a more general equation-of-state parameter \cite{DES:2021wwk}.  Similar shifts arise from different assumptions about still-undetermined neutrino masses.  Although these shifts are small, at the $1\sigma$ level, they can change a result from the $>3\sigma$ threshold to a $<3\sigma$ result.  At this level, conclusions can also depend on the interpretation of the statistics.
For example, Refs.~\cite{Ivanov:2020ril,Hill:2020osr,DAmico:2020ods} argued that current data favored $\Lambda$CDM over EDE at the
$\gtrsim3\sigma$ level, while others \cite{Smith:2020rxx,Herold:2022iib}
warn that the conclusions may reflect the choice of priors.

Perhaps the most intriguing results at the time of writing are reported in Refs.~\cite{Hill:2021yec,Poulin:2021bjr}.  They find that new measurements of small-scale polarization from ACT
Data Release 4 \cite{ACT:2020gnv} favor EDE over
$\Lambda$CDM at the $\gtrsim3 \sigma$ level.  The results should
be considered as provisional given some possible inconsistencies
between Planck and ACT measurements.  Still, the results, based
on a fraction of {\it current} polarization data suggest that
forthcoming experiments, like the Simons Observatory and CMB-S4 (if not ACT/SPT data obtained even sooner), should distinguish EDE from $\Lambda$CDM with high statistical significance. 

Perhaps the most important takeaway is that four years later, with several new data sets (especially for small-scale CMB fluctuations, which were anticipated \cite{Poulin:2018cxd} to provide the most stringent tests of EDE), EDE remains consistent and even possibly favored over $\Lambda$CDM.  We refer readers interested in a more detailed discussion of the statistical techniques, tests, and data sets to Ref.~\cite{Schoneberg:2021qvd}, and to Ref.~\cite{DiValentino:2021izs} for a comprehensive tour of EDE models.

\section{CONCLUSIONS}

The discrepancy between local determinations of the cosmic expansion rate based on distance and redshift measurements and the expansion rate inferred from CMB data and galaxy clustering has over time become statistically more significant with new data and simultaneously survived careful scrutiny of the relevant measurements and analyses.  This Hubble tension is not solved by any quick fix to the standard cosmological model.  If it were the other way (a lower local $H_0$), it could be accommodated with the type of models for (late-time) dark energy that have been considered for 20 years.  Analogous cosmological tensions in the past have yielded to new insights on stellar populations (to explain Hubble's anomalously large initial expansion rate) and fundamental physics (the 1990s discrepancy between the Hubble constant and age of the Universe, explained ultimately by the discovery of the cosmological constant).  Although it remains to be seen how the current Hubble tension will be resolved, it is likely to provide profound new insights into astrophysics or physics.

The most promising new-physics explanation for the Hubble tension is some new early-Universe physics that decreases the sound horizon.  The most popular playground for such ideas has been early dark energy, which reduces the sound horizon by increasing the expansion rate, but there are other models that involve, for example, modified gravity or changes to the primordial-plasma physics.  Any such model, though, is highly constrained by the data---the model must preserve the excellent agreement between disparate and precise data sets and the canonical $\Lambda$CDM model.  Still, there are models that work and, moreover, remain viable even after comparison with several precise new data sets.  Unlike the last (1990s) Hubble tension, which was resolved ultimately by one number---the cosmological constant---the specification of EDE models is more complicated and thus not as easily digested by theorists.  Still, it is up to data to decide, not our prejudices.

Fortunately, the next steps in exploring the Hubble tension are clear.  Moreover, the required observational infrastructure is already in place, as it coincides largely with that assembled to study (late-Universe) dark energy and inflation.  Ultimately, we must continue to explore astrophysical and measurement uncertainties.  As we have learned over and over in cosmology, there is no single bullet---robust conclusions are only reached with multiple observational avenues and a tightly knit web of calibrations, cross-calibrations, and consistency checks.

\begin{summary}[SUMMARY POINTS]
\begin{enumerate}
\item  The values of the Hubble constant
inferred from CMB measurements and galaxy surveys disagree at
the $\gtrsim 5\sigma$ level with those obtained from
measurements of distances and redshifts in the local Universe.
\item The discrepancy has not yielded to any
simple explanations in terms of systematic effects, despite
considerable scrutiny of the CMB and local measurements.
\item  Of the many models with new physics
explored to explain this Hubble tension, those that involve
modifications to early-Universe dynamics seem best able to
satisfy the panoply of cosmological constraints.
\item Although there have been some hints of
EDE in recent analyses, current CMB data are not yet precise and
robust enough to distinguish EDE models from the standard
$\Lambda$CDM model.
\end{enumerate}
\end{summary}

\begin{issues}[FUTURE ISSUES]
\begin{enumerate}
\item Improving local measurements are needed to refine the Hubble constant while maintaining control of systematic errors with $\sim$1\% a target goal and independent tests at $\sim$ 3\% precision providing valuable crosschecks.  Greater specificity is needed to describe any systematic errors that would evade present detection and impact multiple, independent measures.  
\item  Forthcoming experiments that map more precisely the CMB
polarization on smaller angular scales will be required to test
EDE and other new-physics models for the Hubble tension.
\item If future measurements favor EDE models over the standard
$\Lambda$CDM model, it will be important to understand more
deeply the nature of the new physics that provides EDE-like
dynamics in the early Universe and to explore other times when DE may have affected dynamics of the Universe.
\item It will also be important to think about laboratory, or
other non-cosmological, tests of any such new physics.
\end{enumerate}
\end{issues}

\section*{DISCLOSURE STATEMENT}
The authors are not aware of any affiliations, memberships,
funding, or financial holdings that might be perceived as
affecting the objectivity of this review.

\section*{ACKNOWLEDGMENTS}
We thank Foteini with help producing Figure 2 and Jiaxi Wu for help with Figure 4.  We thank Licia Verde with comments on an earlier draft.  MK was supported by NSF Grant No.\ 2112699 and the Simons Foundation.

\bibliography{edearrefs}

\begin{thebibliography}{147}
\expandafter\ifx\csname natexlab\endcsname\relax\def\natexlab#1{#1}\fi
\expandafter\ifx\csname url\endcsname\relax
  \def\url#1{{\tt #1}}\fi
  \def\apj{{Astrophys.\ J.}}
  \def\apjl{{Astrophys.\ J.\ Lett.}}
  \def\araa{{Ann.\ Rev.\ Astron.\ Astrophys.}}
  \def\mnras{{Mon.\ Not.\ R.\ Astron.\ Soc.}}
  \def\aj{{Astron.\ J.}}

\bibitem[Asimov(1994)]{asimov}
Isaac Asimov.
\newblock {\em Asimov's Chronology of Science \& Discovery: Updated and
  Illustrate}.
\newblock Harper-Collins, New York, 1994.

\bibitem[Freedman et~al.(2001)]{Freedman:2001}
W.~L. Freedman et~al.
\newblock {Final results from the Hubble Space Telescope key project to measure
  the Hubble constant}.
\newblock {\em Astrophys. J.}, 553:\penalty0 47--72, 2001, astro-ph/0012376.

\bibitem[{Freedman} et~al.(2012){Freedman}, {Madore}, {Scowcroft}, {Burns},
  {Monson}, {Persson}, {Seibert}, and {Rigby}]{Freedman:2012}
Wendy~L. {Freedman}, Barry~F. {Madore}, Victoria {Scowcroft}, Chris {Burns},
  Andy {Monson}, S.~Eric {Persson}, Mark {Seibert}, and Jane {Rigby}.
\newblock {Carnegie Hubble Program: A Mid-infrared Calibration of the Hubble
  Constant}.
\newblock {\em \apj}, 758\penalty0 (1):\penalty0 24, October 2012, 1208.3281.

\bibitem[Aghanim et~al.(2020)]{Planck:2018vyg}
N.~Aghanim et~al.
\newblock {Planck 2018 results. VI. Cosmological parameters}.
\newblock {\em Astron. Astrophys.}, 641:\penalty0 A6, 2020, 1807.06209.
\newblock [Erratum: Astron.Astrophys. 652, C4 (2021)].

\bibitem[Alam et~al.(2021)]{eBOSS:2020yzd}
Shadab Alam et~al.
\newblock {Completed SDSS-IV extended Baryon Oscillation Spectroscopic Survey:
  Cosmological implications from two decades of spectroscopic surveys at the
  Apache Point Observatory}.
\newblock {\em Phys. Rev. D}, 103\penalty0 (8):\penalty0 083533, 2021,
  2007.08991.

\bibitem[Abbott et~al.(2022)]{DES:2021wwk}
T.~M.~C. Abbott et~al.
\newblock {Dark Energy Survey Year 3 results: Cosmological constraints from
  galaxy clustering and weak lensing}.
\newblock {\em Phys. Rev. D}, 105\penalty0 (2):\penalty0 023520, 2022,
  2105.13549.

\bibitem[Riess et~al.(2011)Riess, Macri, Casertano, Lampeitl, Ferguson,
  Filippenko, Jha, Li, and Chornock]{Riess:2011yx}
Adam~G. Riess, Lucas Macri, Stefano Casertano, Hubert Lampeitl, Henry~C.
  Ferguson, Alexei~V. Filippenko, Saurabh~W. Jha, Weidong Li, and Ryan
  Chornock.
\newblock {A 3\% Solution: Determination of the Hubble Constant with the Hubble
  Space Telescope and Wide Field Camera 3}.
\newblock {\em Astrophys. J.}, 730:\penalty0 119, 2011, 1103.2976.
\newblock [Erratum: Astrophys.J. 732, 129 (2011)].

\bibitem[Riess et~al.(2016)]{Riess:2016jrr}
Adam~G. Riess et~al.
\newblock {A 2.4\% Determination of the Local Value of the Hubble Constant}.
\newblock {\em Astrophys. J.}, 826\penalty0 (1):\penalty0 56, 2016, 1604.01424.

\bibitem[Riess et~al.(2022{\natexlab{a}})]{Riess:2021jrx}
Adam~G. Riess et~al.
\newblock {A Comprehensive Measurement of the Local Value of the Hubble
  Constant with 1 km/sec/MpcUncertainty from the Hubble Space
  Telescope and the SH0ES Team}.
\newblock {\em Astrophys. J. Lett.}, 934\penalty0 (1):\penalty0 L7,
  2022{\natexlab{a}}, 2112.04510.

\bibitem[Perlmutter et~al.(1999)]{SupernovaCosmologyProject:1998vns}
S.~Perlmutter et~al.
\newblock {Measurements of $\Omega$ and $\Lambda$ from 42 high redshift
  supernovae}.
\newblock {\em Astrophys. J.}, 517:\penalty0 565--586, 1999, astro-ph/9812133.

\bibitem[Riess et~al.(1998)]{SupernovaSearchTeam:1998fmf}
Adam~G. Riess et~al.
\newblock {Observational evidence from supernovae for an accelerating universe
  and a cosmological constant}.
\newblock {\em Astron. J.}, 116:\penalty0 1009--1038, 1998, astro-ph/9805201.

\bibitem[Abdalla et~al.(2022)]{Abdalla:2022yfr}
Elcio Abdalla et~al.
\newblock {Cosmology intertwined: A review of the particle physics,
  astrophysics, and cosmology associated with the cosmological tensions and
  anomalies}.
\newblock {\em JHEAp}, 34:\penalty0 49--211, 2022, 2203.06142.

\bibitem[Caldwell and Kamionkowski(2009)]{Caldwell:2009ix}
Robert~R. Caldwell and Marc Kamionkowski.
\newblock {The Physics of Cosmic Acceleration}.
\newblock {\em Ann. Rev. Nucl. Part. Sci.}, 59:\penalty0 397--429, 2009,
  0903.0866.

\bibitem[Karwal and Kamionkowski(2016)]{Karwal:2016vyq}
Tanvi Karwal and Marc Kamionkowski.
\newblock {Dark energy at early times, the Hubble parameter, and the string
  axiverse}.
\newblock {\em Phys. Rev. D}, 94\penalty0 (10):\penalty0 103523, 2016,
  1608.01309.

\bibitem[Poulin et~al.(2018{\natexlab{a}})Poulin, Smith, Grin, Karwal, and
  Kamionkowski]{Poulin:2018dzj}
Vivian Poulin, Tristan~L. Smith, Daniel Grin, Tanvi Karwal, and Marc
  Kamionkowski.
\newblock {Cosmological implications of ultralight axionlike fields}.
\newblock {\em Phys. Rev. D}, 98\penalty0 (8):\penalty0 083525,
  2018{\natexlab{a}}, 1806.10608.

\bibitem[Bernal et~al.(2016)Bernal, Verde, and Riess]{Bernal:2016gxb}
Jose~Luis Bernal, Licia Verde, and Adam~G. Riess.
\newblock {The trouble with $H_0$}.
\newblock {\em JCAP}, 10:\penalty0 019, 2016, 1607.05617.

\bibitem[Verde et~al.(2019)Verde, Treu, and Riess]{Verde:2019ivm}
L.~Verde, T.~Treu, and A.~G. Riess.
\newblock {Tensions between the Early and the Late Universe}.
\newblock {\em Nature Astron.}, 3:\penalty0 891, 7 2019, 1907.10625.

\bibitem[Knox and Millea(2020)]{Knox:2019rjx}
Lloyd Knox and Marius Millea.
\newblock {Hubble constant hunter\textquoteright{}s guide}.
\newblock {\em Phys. Rev. D}, 101\penalty0 (4):\penalty0 043533, 2020,
  1908.03663.

\bibitem[Di~Valentino et~al.(2021)Di~Valentino, Mena, Pan, Visinelli, Yang,
  Melchiorri, Mota, Riess, and Silk]{DiValentino:2021izs}
Eleonora Di~Valentino, Olga Mena, Supriya Pan, Luca Visinelli, Weiqiang Yang,
  Alessandro Melchiorri, David~F. Mota, Adam~G. Riess, and Joseph Silk.
\newblock {In the realm of the Hubble tension\textemdash{}a review of
  solutions}.
\newblock {\em Class. Quant. Grav.}, 38\penalty0 (15):\penalty0 153001, 2021,
  2103.01183.

\bibitem[{Shah} et~al.(2021){Shah}, {Lemos}, and {Lahav}]{Shah:2021}
Paul {Shah}, Pablo {Lemos}, and Ofer {Lahav}.
\newblock {A buyer's guide to the Hubble constant}.
\newblock {\em Astron.\ Astrophys.\ Rev.}, 29\penalty0 (1):\penalty0 9, December 2021, 2109.01161.

\bibitem[{Efstathiou}(2021)]{Efstathiou:2021}
George {Efstathiou}.
\newblock {To H$_{0}$ or not to H$_{0}$?}
\newblock {\em \mnras}, 505\penalty0 (3):\penalty0 3866--3872, August 2021,
  2103.08723.

\bibitem[Sch\"oneberg et~al.(2022)Sch\"oneberg, Franco~Abell\'an,
  P\'erez~S\'anchez, Witte, Poulin, and Lesgourgues]{Schoneberg:2021qvd}
Nils Sch\"oneberg, Guillermo Franco~Abell\'an, Andrea P\'erez~S\'anchez,
  Samuel~J. Witte, Vivian Poulin, and Julien Lesgourgues.
\newblock {The H0 Olympics: A fair ranking of proposed models}.
\newblock {\em Phys. Rept.}, 984:\penalty0 1--55, 2022, 2107.10291.

\bibitem[Frieman et~al.(2008)Frieman, Turner, and Huterer]{Frieman:2008sn}
Joshua Frieman, Michael Turner, and Dragan Huterer.
\newblock {Dark Energy and the Accelerating Universe}.
\newblock {\em Ann. Rev. Astron. Astrophys.}, 46:\penalty0 385--432, 2008,
  0803.0982.

\bibitem[Weinberg et~al.(2013)Weinberg, Mortonson, Eisenstein, Hirata, Riess,
  and Rozo]{Weinberg:2013agg}
David~H. Weinberg, Michael~J. Mortonson, Daniel~J. Eisenstein, Christopher
  Hirata, Adam~G. Riess, and Eduardo Rozo.
\newblock {Observational Probes of Cosmic Acceleration}.
\newblock {\em Phys. Rept.}, 530:\penalty0 87--255, 2013, 1201.2434.

\bibitem[Brout et~al.(2022)]{Brout:2022}
Dillon Brout et~al.
\newblock {The Pantheon+ Analysis: Cosmological Constraints}.
\newblock {\em Astrophys. J.}, 938\penalty0 (2):\penalty0 110, 2022,
  2202.04077.

\bibitem[Riess et~al.(2019)Riess, Casertano, Yuan, Macri, and
  Scolnic]{Riess:2019cxk}
Adam~G. Riess, Stefano Casertano, Wenlong Yuan, Lucas~M. Macri, and Dan
  Scolnic.
\newblock {Large Magellanic Cloud Cepheid Standards Provide a 1\% Foundation
  for the Determination of the Hubble Constant and Stronger Evidence for
  Physics beyond LambdaCDM}.
\newblock {\em Astrophys. J.}, 876\penalty0 (1):\penalty0 85, 2019, 1903.07603.

\bibitem[{Christy}(1966)]{1966ARA&A...4..353C}
R.~F. {Christy}.
\newblock {Pulsation Theory}.
\newblock {\em \araa}, 4:\penalty0 353, January 1966.

\bibitem[Pesce et~al.(2020)]{Pesce:2020xfe}
D.~W. Pesce et~al.
\newblock {The Megamaser Cosmology Project. XIII. Combined Hubble constant
  constraints}.
\newblock {\em Astrophys. J. Lett.}, 891\penalty0 (1):\penalty0 L1, 2020,
  2001.09213.

\bibitem[Riess et~al.(2018)]{Riess:2018byc}
Adam~G. Riess et~al.
\newblock {Milky Way Cepheid Standards for Measuring Cosmic Distances and
  Application to Gaia DR2: Implications for the Hubble Constant}.
\newblock {\em Astrophys. J.}, 861\penalty0 (2):\penalty0 126, 2018,
  1804.10655.

\bibitem[Riess et~al.(2021)Riess, Casertano, Yuan, Bowers, Macri, Zinn, and
  Scolnic]{Riess:2020fzl}
Adam~G. Riess, Stefano Casertano, Wenlong Yuan, J.~Bradley Bowers, Lucas Macri,
  Joel~C. Zinn, and Dan Scolnic.
\newblock {Cosmic Distances Calibrated to 1\% Precision with Gaia EDR3
  Parallaxes and Hubble Space Telescope Photometry of 75 Milky Way Cepheids
  Confirm Tension with $\Lambda$CDM}.
\newblock {\em Astrophys. J. Lett.}, 908\penalty0 (1):\penalty0 L6, 2021,
  2012.08534.

\bibitem[Yuan et~al.(2022)Yuan, Riess, Casertano, and Macri]{Yuan:2022edy}
Wenlong Yuan, Adam~G. Riess, Stefano Casertano, and Lucas~M. Macri.
\newblock {A First Look at Cepheids in a SN Ia Host with JWST}.
\newblock 9 2022, 2209.09101.

\bibitem[Scolnic et~al.(2018)]{Scolnic:2018}
D.~M. Scolnic et~al.
\newblock {The Complete Light-curve Sample of Spectroscopically Confirmed SNe
  Ia from Pan-STARRS1 and Cosmological Constraints from the Combined Pantheon
  Sample}.
\newblock {\em Astrophys. J.}, 859\penalty0 (2):\penalty0 101, 2018,
  1710.00845.

\bibitem[Burns et~al.(2018)]{CSP:2018rag}
Christopher~R. Burns et~al.
\newblock {The Carnegie Supernova Project: Absolute Calibration and the Hubble
  Constant}.
\newblock {\em Astrophys. J.}, 869\penalty0 (1):\penalty0 56, 2018, 1809.06381.

\bibitem[Feeney et~al.(2018)Feeney, Mortlock, and Dalmasso]{Feeney:2017sgx}
Stephen~M. Feeney, Daniel~J. Mortlock, and Niccol\`o Dalmasso.
\newblock {Clarifying the Hubble constant tension with a Bayesian hierarchical
  model of the local distance ladder}.
\newblock {\em Mon. Not. Roy. Astron. Soc.}, 476\penalty0 (3):\penalty0
  3861--3882, 2018, 1707.00007.

\bibitem[Javanmardi et~al.(2021)Javanmardi, Merand, Kervella, Breuval,
  Gallenne, Nardetto, Gieren, Pietrzynski, Hocde, and
  Borgniet]{Javanmardi:2021viq}
Behnam Javanmardi, Antoine Merand, Pierre Kervella, Louise Breuval, Alexandre
  Gallenne, Nicolas Nardetto, Wolfgang Gieren, Grzegorz Pietrzynski, Vincent
  Hocde, and Simon Borgniet.
\newblock {Inspecting the Cepheid Distance Ladder: the Hubble Space Telescope
  Distance to the SN Ia Host Galaxy NGC 5584}.
\newblock {\em Astrophys. J.}, 911\penalty0 (1):\penalty0 12, 2021, 2102.12489.

\bibitem[{Breuval} et~al.(2022){Breuval}, {Riess}, and
  {Kervella}]{Breuval:2022}
Louise {Breuval}, Adam~G. {Riess}, and Pierre {Kervella}.
\newblock {An Improved Calibration of the Wavelength Dependence of Metallicity
  on the Cepheid Leavitt law}.
\newblock {\em arXiv e-prints}, page arXiv:2205.06280, May 2022, 2205.06280.

\bibitem[Riess et~al.(2022{\natexlab{b}})Riess, Breuval, Yuan, Casertano,
  \textasciitilde{}Macri, Bowers, Scolnic, Cantat-Gaudin, Anderson, and
  Reyes]{Riess:2022cl}
Adam~G. Riess, Louise Breuval, Wenlong Yuan, Stefano Casertano, Lucas~M.
  \textasciitilde{}Macri, J.~Bradley Bowers, Dan Scolnic, Tristan
  Cantat-Gaudin, Richard~I. Anderson, and Mauricio~Cruz Reyes.
\newblock {Cluster Cepheids with High Precision Gaia Parallaxes, Low Zero-point
  Uncertainties, and Hubble Space Telescope Photometry}.
\newblock {\em Astrophys. J.}, 938\penalty0 (1):\penalty0 36,
  2022{\natexlab{b}}, 2208.01045.

\bibitem[{Hoyt} et~al.(2021){Hoyt}, {Beaton}, {Freedman}, {Jang}, {Lee},
  {Madore}, {Monson}, {Neeley}, {Rich}, and {Seibert}]{Hoyt:2021}
Taylor~J. {Hoyt}, Rachael~L. {Beaton}, Wendy~L. {Freedman}, In~Sung {Jang},
  Myung~Gyoon {Lee}, Barry~F. {Madore}, Andrew~J. {Monson}, Jillian~R.
  {Neeley}, Jeffrey~A. {Rich}, and Mark {Seibert}.
\newblock {The Carnegie Chicago Hubble Program X: Tip of the Red Giant Branch
  Distances to NGC 5643 and NGC 1404}.
\newblock {\em \apj}, 915\penalty0 (1):\penalty0 34, July 2021, 2101.12232.

\bibitem[Freedman et~al.(2019)]{Freedman:2019jwv}
Wendy~L. Freedman et~al.
\newblock {The Carnegie-Chicago Hubble Program. VIII. An Independent
  Determination of the Hubble Constant Based on the Tip of the Red Giant
  Branch}.
\newblock 7 2019, 1907.05922.

\bibitem[{Anand} et~al.(2021){Anand}, {Rizzi}, {Tully}, {Shaya},
  {Karachentsev}, {Makarov}, {Makarova}, {Wu}, {Dolphin}, and
  {Kourkchi}]{2021AJ....162...80A}
Gagandeep~S. {Anand}, Luca {Rizzi}, R.~Brent {Tully}, Edward~J. {Shaya},
  Igor~D. {Karachentsev}, Dmitry~I. {Makarov}, Lidia {Makarova}, Po-Feng {Wu},
  Andrew~E. {Dolphin}, and Ehsan {Kourkchi}.
\newblock {The Extragalactic Distance Database: The Color-Magnitude
  Diagrams/Tip of the Red Giant Branch Distance Catalog}.
\newblock {\em Astron.\ J.}, 162\penalty0 (2):\penalty0 80, August 2021,
  2104.02649.

\bibitem[{Blakeslee} et~al.(2021){Blakeslee}, {Jensen}, {Ma}, {Milne}, and
  {Greene}]{2021ApJ...911...65B}
John~P. {Blakeslee}, Joseph~B. {Jensen}, Chung-Pei {Ma}, Peter~A. {Milne}, and
  Jenny~E. {Greene}.
\newblock {The Hubble Constant from Infrared Surface Brightness Fluctuation
  Distances}.
\newblock {\em Astrophys.\ J.}, 911\penalty0 (1):\penalty0 65, April 2021,
  2101.02221.

\bibitem[{Li} et~al.(2022){Li}, {Casertano}, and {Riess}]{Li22}
Siyang {Li}, Stefano {Casertano}, and Adam~G. {Riess}.
\newblock {A Maximum Likelihood Calibration of the Tip of the Red Giant Branch
  Luminosity from High Latitude Field Giants using Gaia Early Data Release 3
  Parallaxes}.
\newblock {\em arXiv e-prints}, page arXiv:2202.11110, February 2022,
  2202.11110.

\bibitem[{Soltis} et~al.(2021){Soltis}, {Casertano}, and {Riess}]{Soltis21}
John {Soltis}, Stefano {Casertano}, and Adam~G. {Riess}.
\newblock {The Parallax of {\ensuremath{\omega}} Centauri Measured from Gaia
  EDR3 and a Direct, Geometric Calibration of the Tip of the Red Giant Branch
  and the Hubble Constant}.
\newblock {\em \apjl}, 908\penalty0 (1):\penalty0 L5, February 2021,
  2012.09196.

\bibitem[Peterson et~al.(2021)]{Peterson:2021hel}
Erik~R. Peterson et~al.
\newblock {The Pantheon+ Analysis: Evaluating Peculiar Velocity Corrections in
  Cosmological Analyses with Nearby Type Ia Supernovae}.
\newblock 10 2021, 2110.03487.

\bibitem[Brownsberger et~al.(2021)Brownsberger, Brout, Scolnic, Stubbs, and
  Riess]{Brownsberger:2021uue}
Sasha Brownsberger, Dillon Brout, Daniel Scolnic, Christopher~W. Stubbs, and
  Adam~G. Riess.
\newblock {The Pantheon+ Analysis: Dependence of Cosmological Constraints on
  Photometric-Zeropoint Uncertainties of Supernova Surveys}.
\newblock 10 2021, 2110.03486.

\bibitem[Schutz(1986)]{Schutz:1986gp}
Bernard~F. Schutz.
\newblock {Determining the Hubble Constant from Gravitational Wave
  Observations}.
\newblock {\em Nature}, 323:\penalty0 310--311, 1986.

\bibitem[Abbott et~al.(2017{\natexlab{a}})]{LIGOScientific:2017vwq}
B.~P. Abbott et~al.
\newblock {GW170817: Observation of Gravitational Waves from a Binary Neutron
  Star Inspiral}.
\newblock {\em Phys. Rev. Lett.}, 119\penalty0 (16):\penalty0 161101,
  2017{\natexlab{a}}, 1710.05832.

\bibitem[Abbott et~al.(2017{\natexlab{b}})]{LIGOScientific:2017ync}
B.~P. Abbott et~al.
\newblock {Multi-messenger Observations of a Binary Neutron Star Merger}.
\newblock {\em Astrophys. J. Lett.}, 848\penalty0 (2):\penalty0 L12,
  2017{\natexlab{b}}, 1710.05833.

\bibitem[Abbott et~al.(2017{\natexlab{c}})]{LIGOScientific:2017zic}
B.~P. Abbott et~al.
\newblock {Gravitational Waves and Gamma-rays from a Binary Neutron Star
  Merger: GW170817 and GRB 170817A}.
\newblock {\em Astrophys. J. Lett.}, 848\penalty0 (2):\penalty0 L13,
  2017{\natexlab{c}}, 1710.05834.

\bibitem[Abbott et~al.(2017{\natexlab{d}})]{LIGOScientific:2017adf}
B.~P. Abbott et~al.
\newblock {A gravitational-wave standard siren measurement of the Hubble
  constant}.
\newblock {\em Nature}, 551\penalty0 (7678):\penalty0 85--88,
  2017{\natexlab{d}}, 1710.05835.

\bibitem[Chen et~al.(2018)Chen, Fishbach, and Holz]{Chen:2017rfc}
Hsin-Yu Chen, Maya Fishbach, and Daniel~E. Holz.
\newblock {A two per cent Hubble constant measurement from standard sirens
  within five years}.
\newblock {\em Nature}, 562\penalty0 (7728):\penalty0 545--547, 2018,
  1712.06531.

\bibitem[Soares-Santos et~al.(2019)]{DES:2019ccw}
M.~Soares-Santos et~al.
\newblock {First Measurement of the Hubble Constant from a Dark Standard Siren
  using the Dark Energy Survey Galaxies and the LIGO/Virgo
  Binary\textendash{}Black-hole Merger GW170814}.
\newblock {\em Astrophys. J. Lett.}, 876\penalty0 (1):\penalty0 L7, 2019,
  1901.01540.

\bibitem[Huang et~al.(2019)Huang, Riess, Yuan, Macri, Zakamska, Casertano,
  Whitelock, Hoffmann, Filippenko, and Scolnic]{Huang:2019yhh}
Caroline~D. Huang, Adam~G. Riess, Wenlong Yuan, Lucas~M. Macri, Nadia~L.
  Zakamska, Stefano Casertano, Patricia~A. Whitelock, Samantha~L. Hoffmann,
  Alexei~V. Filippenko, and Daniel Scolnic.
\newblock {Hubble Space Telescope Observations of Mira Variables in the Type Ia
  Supernova Host NGC 1559: An Alternative Candle to Measure the Hubble
  Constant}.
\newblock 8 2019, 1908.10883.

\bibitem[{Tonry} and {Schneider}(1988)]{1988AJ.....96..807T}
John {Tonry} and Donald~P. {Schneider}.
\newblock {A New Technique for Measuring Extragalactic Distances}.
\newblock {\em \aj}, 96:\penalty0 807, September 1988.

\bibitem[{Garnavich} et~al.(2022){Garnavich}, {Wood}, {Milne}, {Jensen},
  {Blakeslee}, {Brown}, {Scolnic}, {Rose}, and {Brout}]{Garnavich22}
Peter {Garnavich}, Charlotte~M. {Wood}, Peter {Milne}, Joseph~B. {Jensen},
  John~P. {Blakeslee}, Peter~J. {Brown}, Daniel {Scolnic}, Benjamin {Rose}, and
  Dillon {Brout}.
\newblock {Connecting Infrared Surface Brightness Fluctuation Distances to Type
  Ia Supernova Hosts: Testing the Top Rung of the Distance Ladder}.
\newblock {\em arXiv e-prints}, page arXiv:2204.12060, April 2022, 2204.12060.

\bibitem[EHT()]{EHT} {\url{https://eventhorizontelescope.org}}.


\bibitem[Refsdal(1964)]{Refsdal:1964nw}
S.~Refsdal.
\newblock {On the possibility of determining Hubble's parameter and the masses
  of galaxies from the gravitational lens effect}.
\newblock {\em Mon. Not. Roy. Astron. Soc.}, 128:\penalty0 307, 1964.

\bibitem[{Vanderriest} et~al.(1989){Vanderriest}, {Schneider}, {Herpe},
  {Chevreton}, {Moles}, and {Wlerick}]{1989A&A...215....1V}
C.~{Vanderriest}, J.~{Schneider}, G.~{Herpe}, M.~{Chevreton}, M.~{Moles}, and
  G.~{Wlerick}.
\newblock {The value of the time delay delta T (A,B) for the 'double' quasar
  0957+561 from optical photometric monitoring.}
\newblock {\em Astron.\ Astrophys.}, 215:\penalty0 1--13, May 1989.

\bibitem[Keeton and Kochanek(1997)]{Keeton:1996kf}
C.~R. Keeton and C.~S. Kochanek.
\newblock {Determining the Hubble constant from the gravitational lens
  PG-1115+080}.
\newblock {\em Astrophys. J.}, 487:\penalty0 42, 1997, astro-ph/9611216.

\bibitem[Schechter et~al.(1997)]{Schechter:1996fa}
Paul~L. Schechter et~al.
\newblock {The Quadruple gravitational lens PG1115+080: Time delays and
  models}.
\newblock {\em Astrophys. J. Lett.}, 475:\penalty0 L85--L88, 1997,
  astro-ph/9611051.

\bibitem[Koopmans et~al.(2003)Koopmans, Treu, Fassnacht, Blandford, and
  Surpi]{Koopmans:2003ha}
L.~V.~E. Koopmans, T.~Treu, C.~D. Fassnacht, R.~D. Blandford, and G.~Surpi.
\newblock {The Hubble Constant from the gravitational lens B1608+656}.
\newblock {\em Astrophys. J.}, 599:\penalty0 70--85, 2003, astro-ph/0306216.

\bibitem[Wong et~al.(2020)]{Wong:2019kwg}
Kenneth~C. Wong et~al.
\newblock {H0LiCOW \textendash{} XIII. A 2.4 per cent measurement of H0 from
  lensed quasars: 5.3\ensuremath{\sigma} tension between early- and
  late-Universe probes}.
\newblock {\em Mon. Not. Roy. Astron. Soc.}, 498\penalty0 (1):\penalty0
  1420--1439, 2020, 1907.04869.

\bibitem[{Falco} et~al.(1985){Falco}, {Gorenstein}, and
  {Shapiro}]{1985ApJ...289L...1F}
E.~E. {Falco}, M.~V. {Gorenstein}, and I.~I. {Shapiro}.
\newblock {On model-dependent bounds on H 0 from gravitational images :
  application to Q 0957+561 A, B.}
\newblock {\em Astrophys.\ J.\ Lett.}, 289:\penalty0 L1--L4, February 1985.

\bibitem[Birrer et~al.(2020)]{Birrer:2020tax}
S.~Birrer et~al.
\newblock {TDCOSMO - IV. Hierarchical time-delay cosmography \textendash{}
  joint inference of the Hubble constant and galaxy density profiles}.
\newblock {\em Astron. Astrophys.}, 643:\penalty0 A165, 2020, 2007.02941.

\bibitem[Millon et~al.(2020)]{Millon:2019slk}
M.~Millon et~al.
\newblock {TDCOSMO. I. An exploration of systematic uncertainties in the
  inference of $H_0$ from time-delay cosmography}.
\newblock {\em Astron. Astrophys.}, 639:\penalty0 A101, 2020, 1912.08027.

\bibitem[Bernal et~al.(2021)Bernal, Verde, Jimenez, Kamionkowski, Valcin, and
  Wandelt]{Bernal:2021yli}
Jos\'e~Luis Bernal, Licia Verde, Raul Jimenez, Marc Kamionkowski, David Valcin,
  and Benjamin~D. Wandelt.
\newblock {The trouble beyond $H_0$ and the new cosmic triangles}.
\newblock {\em Phys. Rev. D}, 103\penalty0 (10):\penalty0 103533, 2021,
  2102.05066.

\bibitem[Jimenez and Loeb(2002)]{Jimenez:2001gg}
Raul Jimenez and Abraham Loeb.
\newblock {Constraining cosmological parameters based on relative galaxy ages}.
\newblock {\em Astrophys. J.}, 573:\penalty0 37--42, 2002, astro-ph/0106145.

\bibitem[Stern et~al.(2010)Stern, Jimenez, Verde, Kamionkowski, and
  Stanford]{Stern:2009ep}
Daniel Stern, Raul Jimenez, Licia Verde, Marc Kamionkowski, and S.~Adam
  Stanford.
\newblock {Cosmic Chronometers: Constraining the Equation of State of Dark
  Energy. I: H(z) Measurements}.
\newblock {\em JCAP}, 02:\penalty0 008, 2010, 0907.3149.

\bibitem[Moresco et~al.(2012)]{Moresco:2012jh}
M.~Moresco et~al.
\newblock {Improved constraints on the expansion rate of the Universe up to
  z\textasciitilde{}1.1 from the spectroscopic evolution of cosmic
  chronometers}.
\newblock {\em JCAP}, 08:\penalty0 006, 2012, 1201.3609.

\bibitem[Moresco et~al.(2020)Moresco, Jimenez, Verde, Cimatti, and
  Pozzetti]{Moresco:2020fbm}
Michele Moresco, Raul Jimenez, Licia Verde, Andrea Cimatti, and Lucia Pozzetti.
\newblock {Setting the Stage for Cosmic Chronometers. II. Impact of Stellar
  Population Synthesis Models Systematics and Full Covariance Matrix}.
\newblock {\em Astrophys. J.}, 898\penalty0 (1):\penalty0 82, 2020, 2003.07362.

\bibitem[Borghi et~al.(2022)Borghi, Moresco, Cimatti, Huchet, Quai, and
  Pozzetti]{Borghi:2021zsr}
Nicola Borghi, Michele Moresco, Andrea Cimatti, Alexandre Huchet, Salvatore
  Quai, and Lucia Pozzetti.
\newblock {Toward a Better Understanding of Cosmic Chronometers: Stellar
  Population Properties of Passive Galaxies at Intermediate Redshift}.
\newblock {\em Astrophys. J.}, 927\penalty0 (2):\penalty0 164, 2022,
  2106.14894.

\bibitem[{O'Malley} et~al.(2017){O'Malley}, {Gilligan}, and
  {Chaboyer}]{2017ApJ...838..162O}
Erin~M. {O'Malley}, Christina {Gilligan}, and Brian {Chaboyer}.
\newblock {Absolute Ages and Distances of 22 GCs Using Monte Carlo
  Main-sequence Fitting}.
\newblock {\em \apj}, 838\penalty0 (2):\penalty0 162, April 2017, 1703.01915.

\bibitem[Jimenez et~al.(2019)Jimenez, Cimatti, Verde, Moresco, and
  Wandelt]{Jimenez:2019onw}
Raul Jimenez, Andrea Cimatti, Licia Verde, Michele Moresco, and Benjamin
  Wandelt.
\newblock {The local and distant Universe: stellar ages and $H_0$}.
\newblock {\em JCAP}, 03:\penalty0 043, 2019, 1902.07081.

\bibitem[Jungman et~al.(1996)Jungman, Kamionkowski, Kosowsky, and
  Spergel]{Jungman:1995bz}
Gerard Jungman, Marc Kamionkowski, Arthur Kosowsky, and David~N. Spergel.
\newblock {Cosmological parameter determination with microwave background
  maps}.
\newblock {\em Phys. Rev. D}, 54:\penalty0 1332--1344, 1996, astro-ph/9512139.

\bibitem[Fixsen et~al.(1996)Fixsen, Cheng, Gales, Mather, Shafer, and
  Wright]{Fixsen:1996nj}
D.~J. Fixsen, E.~S. Cheng, J.~M. Gales, John~C. Mather, R.~A. Shafer, and E.~L.
  Wright.
\newblock {The Cosmic Microwave Background spectrum from the full COBE FIRAS
  data set}.
\newblock {\em Astrophys. J.}, 473:\penalty0 576, 1996, astro-ph/9605054.

\bibitem[Dodelson and Turner(1992)]{Dodelson:1992km}
Scott Dodelson and Michael~S. Turner.
\newblock {Nonequilibrium neutrino statistical mechanics in the expanding
  universe}.
\newblock {\em Phys. Rev. D}, 46:\penalty0 3372--3387, 1992.

\bibitem[Jaffe et~al.(2001)]{Boomerang:2000jdg}
Andrew~H. Jaffe et~al.
\newblock {Cosmology from MAXIMA-1, BOOMERANG and COBE / DMR CMB observations}.
\newblock {\em Phys. Rev. Lett.}, 86:\penalty0 3475--3479, 2001,
  astro-ph/0007333.

\bibitem[Bennett et~al.(2013)]{WMAP:2012fli}
C.~L. Bennett et~al.
\newblock {Nine-Year Wilkinson Microwave Anisotropy Probe (WMAP) Observations:
  Final Maps and Results}.
\newblock {\em Astrophys. J. Suppl.}, 208:\penalty0 20, 2013, 1212.5225.

\bibitem[Spergel et~al.(2003)]{WMAP:2003elm}
D.~N. Spergel et~al.
\newblock {First year Wilkinson Microwave Anisotropy Probe (WMAP) observations:
  Determination of cosmological parameters}.
\newblock {\em Astrophys. J. Suppl.}, 148:\penalty0 175--194, 2003,
  astro-ph/0302209.

\bibitem[Aiola et~al.(2020)]{ACT:2020gnv}
Simone Aiola et~al.
\newblock {The Atacama Cosmology Telescope: DR4 Maps and Cosmological
  Parameters}.
\newblock {\em JCAP}, 12:\penalty0 047, 2020, 2007.07288.

\bibitem[Dutcher et~al.(2021)]{SPT-3G:2021eoc}
D.~Dutcher et~al.
\newblock {Measurements of the E-mode polarization and temperature-E-mode
  correlation of the CMB from SPT-3G 2018 data}.
\newblock {\em Phys. Rev. D}, 104\penalty0 (2):\penalty0 022003, 2021,
  2101.01684.

\bibitem[Ross et~al.(2017)]{BOSS:2016apd}
Ashley~J. Ross et~al.
\newblock {The clustering of galaxies in the completed SDSS-III Baryon
  Oscillation Spectroscopic Survey: Observational systematics and baryon
  acoustic oscillations in the correlation function}.
\newblock {\em Mon. Not. Roy. Astron. Soc.}, 464\penalty0 (1):\penalty0
  1168--1191, 2017, 1607.03145.

\bibitem[Brieden et~al.(2021)Brieden, Gil-Mar\'\i{}n, and
  Verde]{Brieden:2021cfg}
Samuel Brieden, H\'ector Gil-Mar\'\i{}n, and Licia Verde.
\newblock {Model-independent versus model-dependent interpretation of the
  SDSS-III BOSS power spectrum: Bridging the divide}.
\newblock {\em Phys. Rev. D}, 104\penalty0 (12):\penalty0 L121301, 2021,
  2106.11931.

\bibitem[Philcox et~al.(2022)Philcox, Farren, Sherwin, Baxter, and
  Brout]{Philcox:2022sgj}
Oliver H.~E. Philcox, Gerrit~S. Farren, Blake~D. Sherwin, Eric~J. Baxter, and
  Dillon~J. Brout.
\newblock {Determining the Hubble Constant without the Sound Horizon: A $3.6\%$
  Constraint on $H_0$ from Galaxy Surveys, CMB Lensing and Supernovae}.
\newblock 4 2022, 2204.02984.

\bibitem[Smith et~al.(2022)Smith, Poulin, and Simon]{Smith:2022iax}
Tristan~L. Smith, Vivian Poulin, and Th\'eo Simon.
\newblock {Assessing the robustness of sound horizon-free determinations of the
  Hubble constant}.
\newblock 8 2022, 2208.12992.

\bibitem[Poulin et~al.(2018{\natexlab{b}})Poulin, Boddy, Bird, and
  Kamionkowski]{Poulin:2018zxs}
Vivian Poulin, Kimberly~K. Boddy, Simeon Bird, and Marc Kamionkowski.
\newblock {Implications of an extended dark energy cosmology with massive
  neutrinos for cosmological tensions}.
\newblock {\em Phys. Rev. D}, 97\penalty0 (12):\penalty0 123504,
  2018{\natexlab{b}}, 1803.02474.

\bibitem[Caldwell(2002)]{Caldwell:1999ew}
R.~R. Caldwell.
\newblock {A Phantom menace?}
\newblock {\em Phys. Lett. B}, 545:\penalty0 23--29, 2002, astro-ph/9908168.

\bibitem[Efstathiou(2021)]{Efstathiou:2021ocp}
George Efstathiou.
\newblock {To H0 or not to H0?}
\newblock {\em Mon. Not. Roy. Astron. Soc.}, 505\penalty0 (3):\penalty0
  3866--3872, 2021, 2103.08723.

\bibitem[Keeley and Shafieloo(2022)]{Keeley:2022ojz}
Ryan~E. Keeley and Arman Shafieloo.
\newblock {Ruling Out New Physics at Low Redshift as a solution to the $H_0$
  Tension}.
\newblock 6 2022, 2206.08440.

\bibitem[Poulin et~al.(2019)Poulin, Smith, Karwal, and
  Kamionkowski]{Poulin:2018cxd}
Vivian Poulin, Tristan~L. Smith, Tanvi Karwal, and Marc Kamionkowski.
\newblock {Early Dark Energy Can Resolve The Hubble Tension}.
\newblock {\em Phys. Rev. Lett.}, 122\penalty0 (22):\penalty0 221301, 2019,
  1811.04083.

\bibitem[Kamionkowski et~al.(2014)Kamionkowski, Pradler, and
  Walker]{Kamionkowski:2014zda}
Marc Kamionkowski, Josef Pradler, and Devin G.~E. Walker.
\newblock {Dark energy from the string axiverse}.
\newblock {\em Phys. Rev. Lett.}, 113\penalty0 (25):\penalty0 251302, 2014,
  1409.0549.

\bibitem[McDonough and Scalisi(2022)]{McDonough:2022pku}
Evan McDonough and Marco Scalisi.
\newblock {Towards Early Dark Energy in String Theory}.
\newblock 8 2022, 2209.00011.

\bibitem[Turner(1983)]{Turner:1983he}
Michael~S. Turner.
\newblock {Coherent Scalar Field Oscillations in an Expanding Universe}.
\newblock {\em Phys. Rev. D}, 28:\penalty0 1243, 1983.

\bibitem[Johnson and Kamionkowski(2008)]{Johnson:2008se}
Matthew~C. Johnson and Marc Kamionkowski.
\newblock {Dynamical and Gravitational Instability of Oscillating-Field Dark
  Energy and Dark Matter}.
\newblock {\em Phys. Rev. D}, 78:\penalty0 063010, 2008, 0805.1748.

\bibitem[Agrawal et~al.(2019)Agrawal, Cyr-Racine, Pinner, and
  Randall]{Agrawal:2019lmo}
Prateek Agrawal, Francis-Yan Cyr-Racine, David Pinner, and Lisa Randall.
\newblock {Rock 'n' Roll Solutions to the Hubble Tension}.
\newblock 4 2019, 1904.01016.

\bibitem[Lin et~al.(2019{\natexlab{a}})Lin, Benevento, Hu, and
  Raveri]{Lin:2019qug}
Meng-Xiang Lin, Giampaolo Benevento, Wayne Hu, and Marco Raveri.
\newblock {Acoustic Dark Energy: Potential Conversion of the Hubble Tension}.
\newblock {\em Phys. Rev. D}, 100\penalty0 (6):\penalty0 063542,
  2019{\natexlab{a}}, 1905.12618.

\bibitem[Hu(1998)]{Hu:1998kj}
Wayne Hu.
\newblock {Structure formation with generalized dark matter}.
\newblock {\em Astrophys. J.}, 506:\penalty0 485--494, 1998, astro-ph/9801234.

\bibitem[Bertschinger(1995)]{Bertschinger:1995er}
Edmund Bertschinger.
\newblock {COSMICS: cosmological initial conditions and microwave anisotropy
  codes}.
\newblock 6 1995, astro-ph/9506070.

\bibitem[Seljak and Zaldarriaga(1996)]{Seljak:1996is}
Uros Seljak and Matias Zaldarriaga.
\newblock {A Line of sight integration approach to cosmic microwave background
  anisotropies}.
\newblock {\em Astrophys. J.}, 469:\penalty0 437--444, 1996, astro-ph/9603033.

\bibitem[Lewis et~al.(2000)Lewis, Challinor, and Lasenby]{Lewis:1999bs}
Antony Lewis, Anthony Challinor, and Anthony Lasenby.
\newblock {Efficient computation of CMB anisotropies in closed FRW models}.
\newblock {\em Astrophys. J.}, 538:\penalty0 473--476, 2000, astro-ph/9911177.

\bibitem[Lesgourgues(2011)]{Lesgourgues:2011re}
Julien Lesgourgues.
\newblock {The Cosmic Linear Anisotropy Solving System (CLASS) I: Overview}.
\newblock 4 2011, 1104.2932.

\bibitem[Hlozek et~al.(2018)Hlozek, Marsh, and Grin]{Hlozek:2017zzf}
Ren\'ee Hlozek, David J.~E. Marsh, and Daniel Grin.
\newblock {Using the Full Power of the Cosmic Microwave Background to Probe
  Axion Dark Matter}.
\newblock {\em Mon. Not. Roy. Astron. Soc.}, 476\penalty0 (3):\penalty0
  3063--3085, 2018, 1708.05681.

\bibitem[Sabla and Caldwell(2022)]{Sabla:2022xzj}
Vivian~I. Sabla and Robert~R. Caldwell.
\newblock {The Microphysics of Early Dark Energy}.
\newblock 2 2022, 2202.08291.

\bibitem[Karwal et~al.(2022)Karwal, Raveri, Jain, Khoury, and
  Trodden]{Karwal:2021vpk}
Tanvi Karwal, Marco Raveri, Bhuvnesh Jain, Justin Khoury, and Mark Trodden.
\newblock {Chameleon early dark energy and the Hubble tension}.
\newblock {\em Phys. Rev. D}, 105\penalty0 (6):\penalty0 063535, 2022,
  2106.13290.

\bibitem[Sakstein and Trodden(2020)]{Sakstein:2019fmf}
Jeremy Sakstein and Mark Trodden.
\newblock {Early Dark Energy from Massive Neutrinos as a Natural Resolution of
  the Hubble Tension}.
\newblock {\em Phys. Rev. Lett.}, 124\penalty0 (16):\penalty0 161301, 2020,
  1911.11760.

\bibitem[Berghaus and Karwal(2020)]{Berghaus:2019cls}
Kim~V. Berghaus and Tanvi Karwal.
\newblock {Thermal Friction as a Solution to the Hubble Tension}.
\newblock {\em Phys. Rev. D}, 101\penalty0 (8):\penalty0 083537, 2020,
  1911.06281.

\bibitem[Berghaus and Karwal(2022)]{Berghaus:2022cwf}
Kim~V. Berghaus and Tanvi Karwal.
\newblock {Thermal Friction as a Solution to the Hubble and Large-Scale
  Structure Tensions}.
\newblock 4 2022, 2204.09133.

\bibitem[Aloni et~al.(2022)Aloni, Berlin, Joseph, Schmaltz, and
  Weiner]{Aloni:2021eaq}
Daniel Aloni, Asher Berlin, Melissa Joseph, Martin Schmaltz, and Neal Weiner.
\newblock {A Step in understanding the Hubble tension}.
\newblock {\em Phys. Rev. D}, 105\penalty0 (12):\penalty0 123516, 2022,
  2111.00014.

\bibitem[Harari and Sikivie(1992)]{Harari:1992ea}
Diego Harari and Pierre Sikivie.
\newblock {Effects of a Nambu-Goldstone boson on the polarization of radio
  galaxies and the cosmic microwave background}.
\newblock {\em Phys. Lett. B}, 289:\penalty0 67--72, 1992.

\bibitem[Carroll et~al.(1990)Carroll, Field, and Jackiw]{Carroll:1989vb}
Sean~M. Carroll, George~B. Field, and Roman Jackiw.
\newblock {Limits on a Lorentz and Parity Violating Modification of
  Electrodynamics}.
\newblock {\em Phys. Rev. D}, 41:\penalty0 1231, 1990.

\bibitem[Carroll(1998)]{Carroll:1998zi}
Sean~M. Carroll.
\newblock {Quintessence and the rest of the world}.
\newblock {\em Phys. Rev. Lett.}, 81:\penalty0 3067--3070, 1998,
  astro-ph/9806099.

\bibitem[Capparelli et~al.(2020)Capparelli, Caldwell, and
  Melchiorri]{Capparelli:2019rtn}
Ludovico~M. Capparelli, Robert~R. Caldwell, and Alessandro Melchiorri.
\newblock {Cosmic birefringence test of the Hubble tension}.
\newblock {\em Phys. Rev. D}, 101\penalty0 (12):\penalty0 123529, 2020,
  1909.04621.

\bibitem[Murai et~al.(2022)Murai, Naokawa, Namikawa, and
  Komatsu]{Murai:2022zur}
Kai Murai, Fumihiro Naokawa, Toshiya Namikawa, and Eiichiro Komatsu.
\newblock {Isotropic cosmic birefringence from early dark energy}.
\newblock 9 2022, 2209.07804.

\bibitem[Kamionkowski et~al.(1997)Kamionkowski, Kosowsky, and
  Stebbins]{Kamionkowski:1996ks}
Marc Kamionkowski, Arthur Kosowsky, and Albert Stebbins.
\newblock {Statistics of cosmic microwave background polarization}.
\newblock {\em Phys. Rev. D}, 55:\penalty0 7368--7388, 1997, astro-ph/9611125.

\bibitem[Zaldarriaga and Seljak(1997)]{Zaldarriaga:1996xe}
Matias Zaldarriaga and Uros Seljak.
\newblock {An all sky analysis of polarization in the microwave background}.
\newblock {\em Phys. Rev. D}, 55:\penalty0 1830--1840, 1997, astro-ph/9609170.

\bibitem[Lue et~al.(1999)Lue, Wang, and Kamionkowski]{Lue:1998mq}
Arthur Lue, Li-Min Wang, and Marc Kamionkowski.
\newblock {Cosmological signature of new parity violating interactions}.
\newblock {\em Phys. Rev. Lett.}, 83:\penalty0 1506--1509, 1999,
  astro-ph/9812088.

\bibitem[Lepora(1998)]{Lepora:1998ix}
Nathan~F. Lepora.
\newblock {Cosmological birefringence and the microwave background}.
\newblock 12 1998, gr-qc/9812077.

\bibitem[Liu et~al.(2006)Liu, Lee, and Ng]{Liu:2006uh}
Guo-Chin Liu, Seokcheon Lee, and Kin-Wang Ng.
\newblock {Effect on cosmic microwave background polarization of coupling of
  quintessence to pseudoscalar formed from the electromagnetic field and its
  dual}.
\newblock {\em Phys. Rev. Lett.}, 97:\penalty0 161303, 2006, astro-ph/0606248.

\bibitem[Hotinli et~al.(2022)Hotinli, Holder, Johnson, and
  Kamionkowski]{Hotinli:2022wbk}
Selim~C. Hotinli, Gilbert~P. Holder, Matthew~C. Johnson, and Marc Kamionkowski.
\newblock {Cosmology from the kinetic polarized Sunyaev Zel'dovich effect}.
\newblock 4 2022, 2204.12503.

\bibitem[Lee et~al.(2022)Lee, Hotinli, and Kamionkowski]{Lee:2022udm}
Nanoom Lee, Selim~C. Hotinli, and Marc Kamionkowski.
\newblock {Probing Cosmic Birefringence with Polarized Sunyaev Zel'dovich
  Tomography}.
\newblock 7 2022, 2207.05687.

\bibitem[Smith et~al.(2020)Smith, Poulin, and Amin]{Smith:2019ihp}
Tristan~L. Smith, Vivian Poulin, and Mustafa~A. Amin.
\newblock {Oscillating scalar fields and the Hubble tension: a resolution with
  novel signatures}.
\newblock {\em Phys. Rev. D}, 101\penalty0 (6):\penalty0 063523, 2020,
  1908.06995.

\bibitem[Jeong and Kamionkowski(2020)]{Jeong:2019zaz}
Donghui Jeong and Marc Kamionkowski.
\newblock {Gravitational waves, CMB polarization, and the Hubble tension}.
\newblock {\em Phys. Rev. Lett.}, 124\penalty0 (4):\penalty0 041301, 2020,
  1908.06100.

\bibitem[Griest(2002)]{Griest:2002cu}
Kim Griest.
\newblock {Toward a possible solution to the cosmic coincidence problem}.
\newblock {\em Phys. Rev. D}, 66:\penalty0 123501, 2002, astro-ph/0202052.

\bibitem[Dodelson et~al.(2000)Dodelson, Kaplinghat, and
  Stewart]{Dodelson:2000jtt}
Scott Dodelson, Manoj Kaplinghat, and Ewan Stewart.
\newblock {Solving the Coincidence Problem : Tracking Oscillating Energy}.
\newblock {\em Phys. Rev. Lett.}, 85:\penalty0 5276--5279, 2000,
  astro-ph/0002360.

\bibitem[Sabla and Caldwell(2021)]{Sabla:2021nfy}
Vivian~I. Sabla and Robert~R. Caldwell.
\newblock {No $H_0$ assistance from assisted quintessence}.
\newblock {\em Phys. Rev. D}, 103\penalty0 (10):\penalty0 103506, 2021,
  2103.04999.

\bibitem[Hill and Baxter(2018)]{Hill:2018lfx}
J.~Colin Hill and Eric~J. Baxter.
\newblock {Can Early Dark Energy Explain EDGES?}
\newblock {\em JCAP}, 08:\penalty0 037, 2018, 1803.07555.

\bibitem[Bowman et~al.(2018)Bowman, Rogers, Monsalve, Mozdzen, and
  Mahesh]{Bowman:2018yin}
Judd~D. Bowman, Alan E.~E. Rogers, Raul~A. Monsalve, Thomas~J. Mozdzen, and
  Nivedita Mahesh.
\newblock {An absorption profile centred at 78 megahertz in the sky-averaged
  spectrum}.
\newblock {\em Nature}, 555\penalty0 (7694):\penalty0 67--70, 2018, 1810.05912.

\bibitem[Mu\~noz(2019{\natexlab{a}})]{Munoz:2019fkt}
Julian~B. Mu\~noz.
\newblock {Standard Ruler at Cosmic Dawn}.
\newblock {\em Phys. Rev. Lett.}, 123\penalty0 (13):\penalty0 131301,
  2019{\natexlab{a}}, 1904.07868.

\bibitem[Mu\~noz(2019{\natexlab{b}})]{Munoz:2019rhi}
Julian~B. Mu\~noz.
\newblock {Robust Velocity-induced Acoustic Oscillations at Cosmic Dawn}.
\newblock {\em Phys. Rev. D}, 100\penalty0 (6):\penalty0 063538,
  2019{\natexlab{b}}, 1904.07881.

\bibitem[Sarkar and Kovetz(2022)]{Sarkar:2022mdz}
Debanjan Sarkar and Ely~D. Kovetz.
\newblock {Measuring the cosmic expansion rate using 21-cm velocity acoustic
  oscillations}.
\newblock 10 2022, 2210.16853.

\bibitem[Lin et~al.(2019{\natexlab{b}})Lin, Raveri, and Hu]{Lin:2018nxe}
Meng-Xiang Lin, Marco Raveri, and Wayne Hu.
\newblock {Phenomenology of Modified Gravity at Recombination}.
\newblock {\em Phys. Rev. D}, 99\penalty0 (4):\penalty0 043514,
  2019{\natexlab{b}}, 1810.02333.

\bibitem[Braglia et~al.(2021)Braglia, Ballardini, Finelli, and
  Koyama]{Braglia:2020auw}
Matteo Braglia, Mario Ballardini, Fabio Finelli, and Kazuya Koyama.
\newblock {Early modified gravity in light of the $H_0$ tension and LSS data}.
\newblock {\em Phys. Rev. D}, 103\penalty0 (4):\penalty0 043528, 2021,
  2011.12934.

\bibitem[Braglia et~al.(2020)Braglia, Ballardini, Emond, Finelli, Gumrukcuoglu,
  Koyama, and Paoletti]{Braglia:2020iik}
Matteo Braglia, Mario Ballardini, William~T. Emond, Fabio Finelli, A.~Emir
  Gumrukcuoglu, Kazuya Koyama, and Daniela Paoletti.
\newblock {Larger value for $H_0$ by an evolving gravitational constant}.
\newblock {\em Phys. Rev. D}, 102\penalty0 (2):\penalty0 023529, 2020,
  2004.11161.

\bibitem[Ballesteros et~al.(2020)Ballesteros, Notari, and
  Rompineve]{Ballesteros:2020sik}
Guillermo Ballesteros, Alessio Notari, and Fabrizio Rompineve.
\newblock {The $H_0$ tension: $\Delta G_N$ vs. $\Delta N_{\rm eff}$}.
\newblock {\em JCAP}, 11:\penalty0 024, 2020, 2004.05049.

\bibitem[Ballardini et~al.(2020)Ballardini, Braglia, Finelli, Paoletti,
  Starobinsky, and Umilt\`a]{Ballardini:2020iws}
Mario Ballardini, Matteo Braglia, Fabio Finelli, Daniela Paoletti, Alexei~A.
  Starobinsky, and Caterina Umilt\`a.
\newblock {Scalar-tensor theories of gravity, neutrino physics, and the $H_0$
  tension}.
\newblock {\em JCAP}, 10:\penalty0 044, 2020, 2004.14349.

\bibitem[Zumalacarregui(2020)]{Zumalacarregui:2020cjh}
Miguel Zumalacarregui.
\newblock {Gravity in the Era of Equality: Towards solutions to the Hubble
  problem without fine-tuned initial conditions}.
\newblock {\em Phys. Rev. D}, 102\penalty0 (2):\penalty0 023523, 2020,
  2003.06396.

\bibitem[Abadi and Kovetz(2021)]{Abadi:2020hbr}
Tal Abadi and Ely~D. Kovetz.
\newblock {Can conformally coupled modified gravity solve the Hubble tension?}
\newblock {\em Phys. Rev. D}, 103\penalty0 (2):\penalty0 023530, 2021,
  2011.13853.

\bibitem[Kreisch et~al.(2020)Kreisch, Cyr-Racine, and Dor\'e]{Kreisch:2019yzn}
Christina~D. Kreisch, Francis-Yan Cyr-Racine, and Olivier Dor\'e.
\newblock {Neutrino puzzle: Anomalies, interactions, and cosmological
  tensions}.
\newblock {\em Phys. Rev. D}, 101\penalty0 (12):\penalty0 123505, 2020,
  1902.00534.

\bibitem[Blinov et~al.(2019)Blinov, Kelly, Krnjaic, and
  McDermott]{Blinov:2019gcj}
Nikita Blinov, Kevin~James Kelly, Gordan~Z Krnjaic, and Samuel~D McDermott.
\newblock {Constraining the Self-Interacting Neutrino Interpretation of the
  Hubble Tension}.
\newblock {\em Phys. Rev. Lett.}, 123\penalty0 (19):\penalty0 191102, 2019,
  1905.02727.

\bibitem[Jedamzik and Pogosian(2020)]{Jedamzik:2020krr}
Karsten Jedamzik and Levon Pogosian.
\newblock {Relieving the Hubble tension with primordial magnetic fields}.
\newblock {\em Phys. Rev. Lett.}, 125\penalty0 (18):\penalty0 181302, 2020,
  2004.09487.

\bibitem[Ivanov et~al.(2020)Ivanov, McDonough, Hill, Simonovi\'c, Toomey,
  Alexander, and Zaldarriaga]{Ivanov:2020ril}
Mikhail~M. Ivanov, Evan McDonough, J.~Colin Hill, Marko Simonovi\'c, Michael~W.
  Toomey, Stephon Alexander, and Matias Zaldarriaga.
\newblock {Constraining Early Dark Energy with Large-Scale Structure}.
\newblock {\em Phys. Rev. D}, 102\penalty0 (10):\penalty0 103502, 2020,
  2006.11235.

\bibitem[Hill et~al.(2020)Hill, McDonough, Toomey, and Alexander]{Hill:2020osr}
J.~Colin Hill, Evan McDonough, Michael~W. Toomey, and Stephon Alexander.
\newblock {Early dark energy does not restore cosmological concordance}.
\newblock {\em Phys. Rev. D}, 102\penalty0 (4):\penalty0 043507, 2020,
  2003.07355.

\bibitem[D'Amico et~al.(2021)D'Amico, Senatore, Zhang, and
  Zheng]{DAmico:2020ods}
Guido D'Amico, Leonardo Senatore, Pierre Zhang, and Henry Zheng.
\newblock {The Hubble Tension in Light of the Full-Shape Analysis of
  Large-Scale Structure Data}.
\newblock {\em JCAP}, 05:\penalty0 072, 2021, 2006.12420.

\bibitem[Smith et~al.(2021)Smith, Poulin, Bernal, Boddy, Kamionkowski, and
  Murgia]{Smith:2020rxx}
Tristan~L. Smith, Vivian Poulin, Jos\'e~Luis Bernal, Kimberly~K. Boddy, Marc
  Kamionkowski, and Riccardo Murgia.
\newblock {Early dark energy is not excluded by current large-scale structure
  data}.
\newblock {\em Phys. Rev. D}, 103\penalty0 (12):\penalty0 123542, 2021,
  2009.10740.

\bibitem[Herold and Ferreira(2022)]{Herold:2022iib}
Laura Herold and Elisa G.~M. Ferreira.
\newblock {Resolving the Hubble tension with Early Dark Energy}.
\newblock 10 2022, 2210.16296.

\bibitem[Hill et~al.(2022)]{Hill:2021yec}
J.~Colin Hill et~al.
\newblock {Atacama Cosmology Telescope: Constraints on prerecombination early
  dark energy}.
\newblock {\em Phys. Rev. D}, 105\penalty0 (12):\penalty0 123536, 2022,
  2109.04451.

\bibitem[Poulin et~al.(2021)Poulin, Smith, and Bartlett]{Poulin:2021bjr}
Vivian Poulin, Tristan~L. Smith, and Alexa Bartlett.
\newblock {Dark energy at early times and ACT data: A larger Hubble constant
  without late-time priors}.
\newblock {\em Phys. Rev. D}, 104\penalty0 (12):\penalty0 123550, 2021,
  2109.06229.

\end{thebibliography}
\bibliographystyle{hunsrtnat}

\end{document}